 \documentclass[11pt]{article}
\usepackage{graphicx}
\usepackage{epsfig}
\usepackage{amsfonts}

\usepackage{amsmath}

\newcommand{\beq}{\begin{equation}}
\newcommand{\eeq}{\end{equation}}
\newcommand{\beqs}{\begin{eqnarray}}
\newcommand{\eeqs}{\end{eqnarray}}
\newcommand{\ra}{\rightarrow}

\newcommand{\vphi}{\varphi}

\newcommand{\beqa}{\begin{eqnarray}}
\newcommand{\eeqa}{\end{eqnarray}}
\newcommand{\beqar}{\begin{eqnarray*}}
\newcommand{\eeqar}{\end{eqnarray*}}

\newcommand{\tr}{\mbox{tr}}

\newcommand{\ee}{\end{equation}}
\newcommand{\eea}{\end{eqnarray}}
\newcommand{\be}{\begin{equation}}
\newcommand{\bea}{\begin{eqnarray}}

\def\theequation{\arabic{equation}}

\newcommand{\insertplot}[5]{\begin{figure}
 
\hfill\hbox to 0.05in{\vbox to #5in{\vfill
 
\inputplot{#1}{#4}{#5}}\hfill}
 \hfill\vspace{-.1in}
 \caption{#2}\label{#3}
 \end{figure}}
 
\newcommand{\inputplot}[3]{
 \special{ps: 
plotfile #1}
}
\newcounter{fig}

\begin{document}
\title{Black strings with negative cosmological constant: 
\\inclusion of electric charge 
and rotation}
\author{{\large Yves Brihaye }$^{\dagger}$,
{\large Eugen Radu}$^{\dagger \star}$ 
and {\large Cristian Stelea}$^{\ddagger \star}$ 
\\ \\
$^{\dagger}${\small Facult\'e des Sciences, Universit\'e de Mons-Hainaut,
B-7000 Mons, Belgium } \\
$^{\dagger \star}${\small Department of  Mathematical Physics,
National University of Ireland Maynooth, Ireland}\\
$^{\ddagger \star}${\small Department of Physics and Astronomy, University of Waterloo,
Ontario N2L 3G1, Canada}
} 
 
\maketitle


\medskip
\begin{abstract}
We generalize the vacuum static black string solutions of Einstein's 
equations with negative cosmological constant
recently discussed in literature,
by including an electromagnetic field. 
These higher-dimensional  configurations have no dependence on 
the `compact' extra dimension, 
while their boundary topology is the product of time
and $S^{d-3}\times S^1$ or  $H^{d-3}\times S^1$.
 Rotating generalizations of the even dimensional 
black string configurations are considered as well.
Different from the static, neutral case, no regular limit is found 
for a vanishing event horizon radius.
We explore numerically the general properties of such 
classical solutions and, using a counterterm prescription, we compute their 
conserved charges and discuss their thermodynamics.
We find that the
thermodynamics of the  black strings  follows the pattern of the 
corresponding black hole solutions in AdS backgrounds.
\end{abstract}
 
 
\section{Introduction}  
Over the past decade there has a been a considerable interest 
in the physics of black holes in anti-de Sitter (AdS) backgrounds. 
This interest is mainly motivated by the proposed correspondence 
between physical effects associated with gravitating fields 
propagating in an AdS arena and those of a conformal field 
theory (CFT) on its boundary \cite{Maldacena:1997re,Witten:1998qj}. 
In light of this AdS/CFT correspondence, asymptotically AdS 
 black hole solutions would offer the possibility of studying some 
aspects of the nonperturbative structure of certain quantum
field theories. 
For example, the  Schwarzschild-AdS$_5$ Hawking-Page phase 
transition \cite{Hawking:1982dh} is interpreted as a thermal phase 
transition from a confining to a deconfining phase in the dual 
$D = 4$, ${\cal N} = 4$ super Yang-Mills theory \cite{Witten:1998zw}, 
while the phase structure of Reissner-Nordstr\"om-AdS (RNAdS) 
black holes, which resembles that of a van der Waals-Maxwell 
liquid-gas system is related to the physics of a class of field 
theories coupled to a background current \cite{Chamblin:1999tk}.

In view of this duality,
it is  of interest to find new solutions of Einstein equations 
with negative cosmological constant and study their physics,
trying to relate it to the physics of the boundary CFT.
Both Schwarzschild- and RNAdS black hole solutions in $d$ dimensions
have an event 
horizon of topology $S^{d-2}$,  
which matches the $S^{d-2}$ topology of the spacelike infinity.
In \cite{Copsey:2006br} there were presented both analytical and numerical arguments for the 
existence of a different type of $d=5$ configurations, 
with an event horizon topology $S^{2}\times S^1$. These configurations have no 
dependence on the `compact' extra dimension, 
and their conformal boundary is the product of 
time and $S^{2}\times S^1$  (note that due to the 
non-linearity of the field equations, it was impossible to 
find the exact solutions with such properties).
These solutions have been generalized to higher 
dimensions $d\geq5$ in \cite{ Mann:2006yi},  
configurations with an event horizon topology 
$H^{d-3}\times S^1$ being considered as well. 
Black objects with event horizon topology 
$S^{d-3}\times S^1$ matching that of the 
spacelike infinity are familiar from the
 $\Lambda=0$ physics and they are usually 
called black strings \cite{bs}. 
The solutions in \cite{Copsey:2006br, Mann:2006yi}
present many similar properties with the  $\Lambda=0$ case,
and are naturally interpreted as the AdS counterparts of these 
configurations\footnote{Note that the 
uniform solutions in \cite{Copsey:2006br, Mann:2006yi}
are very different from the warped  AdS black 
strings as discussed for instance in \cite{Chamblin:1999by}.}. However, 
different from the $\Lambda=0$ limit, it was found in \cite{ Mann:2006yi} that the AdS black 
string solutions  with an event horizon 
topology $S^{d-3}\times S^1$ have a nontrivial, 
globally regular limit with zero event 
horizon radius.
As argued in \cite{Copsey:2006br, Mann:2006yi}, these solutions
provide the gravity dual of a field theory  on
a $S^{d-3}\times S^1\times S^1$ (or $H^{d-3}\times S^1\times S^1$)
background.

In this paper we consider further generalizations of these black string solutions 
in two different directions. 
Our considerations will be at the classical level.
First, we explore the black string counterparts
of the RNAdS black holes by studying static 
black string solutions 
of the Einstein-Maxwell (EM) gravity
with negative cosmological constant. 
As expected, all vacuum black strings 
admit electrically charged generalizations. 
However, this is not the case for the 
vortex-like structures described in \cite{ Mann:2006yi}, since no regular 
solution is found in the $r_h\to 0$ limit. Certain $\Lambda<0$
charged black string solutions have been 
known for some time in the literature: 
in five-dimensional AdS backgrounds there 
exists an exact solution describing magnetically charged 
black strings  \cite{Chamseddine:1999xk}. However, for these solutions 
the magnetic charge of the black strings depends non-trivially on the cosmological 
constant and their limit (if any) in which the magnetic 
charge is sent to zero in order to recover the uncharged 
black strings in AdS is still unknown. By contrast, the 
solutions discussed in this paper describe electrically 
charged black strings and, furthermore, for our solutions 
there exists a natural zero charge limit in which case we 
recover the previously known neutral black strings. Other 
interesting AdS solutions whose boundary topology is a fibre 
bundle $S^1\times S^1\hookrightarrow S^2$ have been found 
in \cite{mann1} and later generalised to higher dimensions 
in \cite{Lu} (see also the solutions 
in \cite{Astefanesei:2004kn})\footnote{Other locally asymptotically 
AdS geometries with non-trivial boundary geometries and topologies 
can be found in \cite{Astefanesei:2005eq}.}.
 Their charged counterparts have been found in even dimensions only 
and their properties have been discussed in \cite{TNRN}.

The AdS black strings have also rotating generalizations, 
which are discussed in the second part of our paper. Recall that the $\Lambda=0$ 
counterparts of these solutions can be found by taking 
the direct product of a $(d-1)$-dimensional Myers-Perry solutions 
\cite{Myers:1986un} with a circle. Higher dimensional rotating black 
with AdS asymptotics have been found in \cite{Hawking:1998kw, Gibbons:2004uw}.
 However, due to 
the presence of a cosmological constant, these AdS rotating solutions  cannot be uplifted 
to become black string solutions of higher dimensional Einstein 
gravity with a cosmological constant and one has to find such
 solutions by the brute-force solving of Einstein's equations. 
A general  spinning black string solution in $d$ dimensions is characterized 
by $\lfloor(d-1)/2\rfloor$ angular momenta, 
corresponding to independent rotations in the $\lfloor(d-1)/2\rfloor$ 
orthogonal spatial 2-planes, by its mass/energy and its 
tension\footnote{Here $\lfloor x\rfloor$ denotes the 
integer part of the quantity $x$.}. The general 
solution will present a nontrivial dependence on the $(d-4)/2$ 
angular coordinates, which makes the problem difficult to treat 
numerically. However, in the even-dimensional case, the ansatz 
is greatly simplified by taking the ${\it a priori}$ independent 
 $(d-2)/2$ angular momenta to be equal in order to factorize the 
angular dependence \cite{Kunz:2006eh}. This reduces the problem 
to studying the solutions of five differential equations with dependence 
only on the radial variable $r$. 

In this work we examine the general properties of both charged and rotating 
solutions and compute their global charges by using a counterterm prescription.
We discuss the thermodynamics of the charged and 
rotating black string solutions in both the grand-canonical 
and the canonical ensemble. It turns out that the black string thermodynamics resembles 
closely that of the corresponding charged and rotating black 
holes with spherical horizons in AdS backgrounds \cite{Chamblin:1999tk,Hawking:1998kw}.

Our  paper is structured as follows: in the next section we explain 
the model and derive the basic field equations. We also describe 
the computation of the physical quantities of the solutions such as 
their mass-energy, tension, angular momenta and action. The general 
properties of the static charged black string solutions are presented, 
using numerical methods, in Section $3$, while in Section $4$ we present 
the results obtained by numerical calculations in the case of the rotating 
black string solutions. 
The thermodynamical features of the obtained black string 
solutions are discussed in Section $5$. In Section $6$ we present a method to 
derive other solutions locally equivalent with the black string 
geometries and show how to obtain new rotating black hole solutions in 
the lower dimensional EM-Liouville theory obtained by Kaluza-Klein compactification. 
We give our conclusions and remarks in the final section.

\section{The general formalism}

We start with the following action principle in $d$-spacetime dimensions:
\begin{eqnarray}
\label{action}
I_0=\frac{1}{16 \pi G_d}\int_{\mathcal{M}} d^d x \sqrt{-g}
 (R-2 \Lambda-F^2)
-\frac{1}{8 \pi G_d}\int_{\partial\mathcal{M}} d^{d-1} 
x\sqrt{-\gamma}K,
\end{eqnarray}
where  $G_d$ is the gravitational constant in $d$ dimensions, 
$\Lambda=-(d-1)(d-2)/(2 \ell^2)$ is the cosmological constant 
and $F=dA$ is the electromagnetic field strength. Here $\mathcal{M}$ is 
a $d$-dimensional manifold with metric $g_{\mu \nu }$, $K$ is the trace of the extrinsic 
curvature $K_{ab}=-\gamma_{a}^{c}\nabla_{c}n_{b}$ of 
the boundary $\partial M$ with unit normal $n^{a}$ and induced metric $\gamma_{ab}$.

As usual, the classical equations of motion are derived by setting 
the variations of the action (\ref{action}) to zero.  In our case, by varying the above action one obtains the 
 EM system of field equations:
\beqs
G_{\mu\nu}-\frac{(d-1)(d-2)}{2\ell^2}g_{\mu\nu}=8\pi G~T_{\mu\nu},~~~
\nabla_{\mu}F^{\mu\nu}=0,
\label{einstein}
\eeqs
together with a Bianchi identity for the electromagnetic field 
$dF=0$. Here $G_{\mu\nu}$ is the Einstein tensor and 
$T_{\mu\nu}=\frac{1}{4\pi G}\left(F_{\mu\sigma}F^{\sigma}_{\nu}-
\frac{1}{4}F_{\sigma\rho}F^{\sigma\rho}g_{\mu\nu}\right)$ 
is the stress tensor of the electromagnetic field.
 However, in deriving the above field equations we must pay particular attention 
to the boundary condition to be imposed on the 
electromagnetic potential $A$. If one keeps the 
value of the electromagnetic potential fixed on the boundary we obtain directly 
(\ref{einstein}) and the action (\ref{action}) will
 be appropriate for studying the thermodynamics of
 the charged black string using the grand-canonical ensemble. On the other hand, 
if one performs a study of the canonical ensemble with
 fixed charge on the boundary, one has to fix the value 
of $n^aF_{ab}$ on the boundary and the action (\ref{action}) 
will have to be modified accordingly \cite{Hawking:1995ap}:
\beqs
I&=&I_0-\frac{1}{4\pi G_d}\int_{\partial\mathcal{M}} d^{d-1} x\sqrt{-\gamma}F^{ab}n_aA_b.
\label{canaction}
\eeqs
Note that the rotating solutions to be discussed 
in Section $4$ are found for a vanishing 
gauge field $A_\mu=0$.

When evaluated on non-compact solutions of the field equations, 
it turns out that the action (\ref{action}) diverges. 
The general remedy for this situation is to add counterterms, 
\textit{i.e.} coordinate invariant functionals of the 
intrinsic boundary geometry that are specifically designed to cancel 
out the divergences. To regularize the divergences in the gravitational action sector, 
 the following boundary counterterm part is added to the action principle (\ref{action})
\cite{Balasubramanian:1999re,Das:2000cu}:
\begin{eqnarray}
I_{\mathrm{ct}}^0 &=&\frac{1}{8\pi G_d}\int d^{d-1}x\sqrt{-\gamma 
}\left\{ -\frac{d-2}{\ell }-\frac{\ell \mathsf{\Theta }\left( d-4\right) 
}{2(d-3)}\mathsf{R}-\frac{\ell ^{3}\mathsf{\Theta }\left( d-6\right) 
}{2(d-3)^{2}(d-5)}\left(\mathsf{R}_{ab}\mathsf{R}^{ab}-
\frac{d-1}{4(d-2)}\mathsf{R}^{2}\right) 
\right.
\nonumber  
\\
\label{Lagrangianct} 
&&+\frac{\ell ^{5}\mathsf{\Theta }\left( d-8\right) 
}{(d-3)^{3}(d-5)(d-7)}\left( 
\frac{3d-1}{4(d-2)}\mathsf{RR}^{ab}\mathsf{R}_{ab}
-\frac{d^2-1}{16(d-2)^{2}}\mathsf{R}^{3}\right.  \nonumber \\
&&\left. -2\mathsf{R}^{ab}\mathsf{R}^{cd}\mathsf{R}_{acbd}\left. 
-\frac{d-1}{4(d-2)}\nabla _{a}\mathsf{R}\nabla ^{a}\mathsf{R}+\nabla 
^{c}\mathsf{R}^{ab}\nabla _{c}\mathsf{R}_{ab}\right) +...\right\} ,
\end{eqnarray}
where $\mathsf{R}$ and $\mathsf{R}^{ab}$ are the curvature 
and the Ricci tensor associated with the induced metric $\gamma $. 
The series truncates for any fixed dimension, with new terms
 entering at every new even value of $d$, as denoted by the 
step-function ($\mathsf{\Theta }\left( x\right) =1$
 provided $x\geq 0$, and vanishes otherwise).

However, as we shall find in the next two sections, 
given the presence for odd $d$ of $\log(r/\ell)$ terms in the 
asymptotic expansions of the metric functions 
 (with $r$ the radial coordinate), the counterterms 
(\ref{Lagrangianct}) regularise the action for 
 even dimensions only. For odd values of $d$, 
we have to add the following extra terms to (\ref{action}) \cite{Skenderis:2000in}:
\begin{eqnarray}
I_{\mathrm{ct}}^{s} &=&\frac{1}{8\pi G_d}\int d^{d-1}x\sqrt{-\gamma 
}\log(\frac{r}{\ell})\left\{  
\mathsf{\delta }_{d,5}\frac{\ell^3 
}{8}(\frac{1}{3}\mathsf{R}^2-\mathsf{R}_{ab}\mathsf{R}^{ab}
)\right.
\nonumber  
\\
&&-\frac{\ell 
^{5}}{128}\left(\mathsf{RR}^{ab}\mathsf{R}_{ab}
-\frac{3}{25}\mathsf{R}^{3} 
-2\mathsf{R}^{ab}\mathsf{R}^{cd}\mathsf{R}_{acbd}\left. 
-\frac{1}{10}\mathsf{R}^{ab}\nabla _{a}\nabla 
_{b}\mathsf{R}+\mathsf{R}^{ab}\Box \mathsf{R}_{ab}
-\frac{1}{10}\mathsf{R}\Box 
\mathsf{R}\right)\delta_{d,7} +\dots
\right\}.\nonumber
\end{eqnarray}%
Using these counterterms in odd and even dimensions, 
one can construct a divergence-free boundary stress 
tensor from the total action $I=I_0+I_{\mathrm{ct}}^0+
I_{\mathrm{ct}}^s$ by defining a boundary stress-tensor: 
\[
T_{ab}=\frac{2}{\sqrt{-\gamma}}\frac{\delta I}{\delta \gamma^{ab}}. 
\]
Consider now a standard ADM decomposition of the metric on the boundary:
\beqs
\gamma_{ab}dx^adx^b=-N^2dt^2+\sigma_{ij}(dy^i+N^idt)(dy^j+N^jdt),
\label{ADMboundary}
\eeqs
where $N$ and $N^i$ are the lapse function, respectively 
the shift vector, and $y^i$, $i=1,\dots,d-2$ are the intrinsic 
coordinates on a closed surface $\Sigma$ of constant time 
$t$ on the boundary. More generally, we can consider an ADM 
decomposition of the spacetime metric in $d$ dimensions, 
which will give rise to (\ref{ADMboundary}) on the boundary. 
Then a conserved charge 
\begin{equation}
{\frak Q}_{\xi }=\oint_{\Sigma }d^{d-2}y\sqrt{\sigma}u^{a}\xi ^{b}T_{ab},
\label{Mcons}
\end{equation}%
can be associated with the closed surface $\Sigma $ 
(with normal $u^{a}$), provided the boundary geometry 
has an isometry generated by a Killing vector $\xi ^{a}$. 
The conserved mass/energy $M$ is the charge associated 
with the time translation symmetry, with $\xi =\partial /\partial t$. 
Similarly to the $\Lambda=0$ case, there is also a second 
charge associated with the compact $z$ direction, corresponding 
to the black string's tension ${\mathcal T}$. 
For the even-dimensional 
spinning solutions considered in Section $4$ there are also 
$(d-2)/2$ angular momenta $J_i$, representing the charges 
associated to the Killing vectors corresponding to angular directions.

For a charged solution, the electric field with respect to a constant $r$
hypersurface is given by $E^{\mu}=g^{\mu\rho}F_{\rho\nu}n^{\nu}$.
 The electric charge of the charged solutions is computed using Gauss' 
law by evaluating the flux of the electric field at infinity:
\begin{eqnarray} 
\label{Qc}  
Q_e=\frac{1}{4\pi G_d}\oint_{\Sigma }d^{d-2}y\sqrt{\sigma}u^{a}n ^{b}F_{ab}.
\end{eqnarray} 
If $A_{\mu}$ is the electromagnetic potential, then the electric 
potential $\Phi$, measured at infinity with respect to the horizon 
is defined as \cite{potential}:
\beqs
\Phi=A_{\mu}\chi^{\mu}|_{r\ra\infty}-A_{\mu}\chi^{\mu}|_{r=r_h}~,
\eeqs
with $\chi^{\mu}$ a  Killing vector orthogonal to and null on the horizon.

The thermodynamics of the black objects is usually 
studied on the Euclidean section \cite{GibbonsHawking1,Hawking:ig}. 
The static vacuum Lorentzian 
solutions discussed in 
\cite{ Mann:2006yi} extremize also the Euclidean
 action as the analytic continuation in time has no effect at the level
 of the equations of motion. However, this is not the case of 
the solutions discussed in this paper since it 
is not possible to find directly real solutions on the 
Euclidean section by Wick rotating $t\ra i\tau$ the Lorentzian 
configurations. Even if one could accompany the Wick rotation 
with various other analytical continuations of the parameters 
describing the solution, given the numerical nature of these
 solutions, there is no  assurance that the modified metric functions 
will also be solutions of the field equations in Euclidean signature. 
In view of this difficulty one  has to resort to an alternative, 
quasi-Euclidean approach as described in \cite{quasi}\footnote{Note 
also that not all solutions with
Lorentzian signature present reasonable Euclidean counterparts, in which case
one is forced to consider a 'quasi-Euclidean' approach.
The $d=5$ asymptotically flat rotating black ring solutions
provides an interesting example in this sense \cite{Astefanesei:2005ad}.
See also \cite{nutty} for a variant 
of the quasi-Euclidean method as applied to NUT-charged spaces.}.
The idea is to regard the action $I$ used in the 
computation of the partition function as a functional 
over complex metrics that are obtained from the real, 
stationary, Lorentzian metrics by using a transformation 
that mimics the effect of the Wick rotation $t\ra i\tau$. 
 In this approach, the values of 
the extensive variables of the complex 
  metric that extremize the path integral are the same as the values 
 of these variables corresponding to the initial Lorentzian metric. 

If there exists an horizon located at $r=r_h$, 
using the quasi-Euclidean metric we can 
compute as usual the Hawking temperature by identifying 
the $\tau$ coordinate with a certain period found by demanding 
regularity of the metric on the horizon. This fixes the 
temperature as the inverse of period $\beta$ of $\tau$. 
It is easy to check that, for the metric forms considered in this paper, 
we obtain the standard relation $T_H=\kappa_H/2\pi$, where:
\begin{eqnarray} 
\label{kappa} 
\kappa_H^2=-\frac{1}{2}\nabla^a \chi^b \nabla_a\chi_b\big|_{r=r_h},
\end{eqnarray}
is the surface gravity, which is constant on the horizon.  

\section{Static charged solutions}

\subsection{The ansatz and equations}

Similarly to the neutral black string metric 
ansatz in \cite{ Mann:2006yi}, we consider 
here the following parametrization of the 
$d$-dimensional line element (with $d \geq 5$):
\begin{eqnarray}
\label{metric} 
ds^2=a(r)dz^2+ \frac{dr^2}{f(r)}+r^2d\Sigma^2_{k,d-3}-b(r)dt^2~,
\end{eqnarray}
where the $(d{-}3)$--dimensional metric $d\Sigma^2_{k,d-3}$ is
\begin{equation}
d\Sigma^2_{k,d-3} =\left\{ \begin{array}{ll}
\vphantom{\sum_{i=1}^{d-3}}
 d\Omega^2_{d-3}& {\rm for}\; k = +1\\
\sum_{i=1}^{d-3} dx_i^2&{\rm for}\; k = 0 \\
\vphantom{\sum_{i=1}^{d-3}}
 d\Xi^2_{d-3} &{\rm for}\; k = -1\ ,
\end{array} \right.
\end{equation}
where $d\Omega^2_{d-3}$ is the unit metric on $S^{d-3}$. By $H^{d-3}$ 
we will understand the $(d{-}3)$--dimensional hyperbolic space, whose unit 
metric  $d\Xi^2_{d-3}$ can be obtained by analytic continuation of 
that on $S^{d-3}$. The direction $z$ is periodic with period $L$.

In the remainder of this paper we shall consider 
an electric ansatz for the electromagnetic field $A=V(r)dt$. 
We can then easily solve the Maxwell equations in (\ref{einstein}) by taking:
\beqs
F^{tr}
=\frac{q}{r^{d-3}}\sqrt{\frac{f(r)}{a(r)b(r)}}~,
\eeqs
where $q$ is an integration constant, to be 
related to the electric charge $Q$. 
We then find the electromagnetic potential $A=V(r)dt$ by a simple integration:
\beqs
\label{Vr}
V(r)&=&\int^r\frac{q}{r^{d-3}}\sqrt{\frac{b(r)}{a(r)f(r)}}dr+\Phi.
\eeqs
Here $\Phi$ is a constant to be fixed later. 

The Einstein-Maxwell equations with a negative cosmological constant imply then
that the metric functions $a(r)$, $b(r)$ 
and $f(r)$ are solutions of the following equations:
\begin{eqnarray}
\label{ep1} 
\nonumber
f'=\frac{2k(d-4)}{r}
+\frac{2(d-1)r}{\ell^2}
-\frac{2(d-4)f}{r}
-f\left(\frac{a'}{a}+\frac{b'}{b} \right)-\frac{4q^2}{(d-2)r^{2d-7}a}~,
\end{eqnarray}
\begin{eqnarray}
\label{ep2} 
b''&=&\frac{(d-3)(d-4)b}{r^2} 
-\frac{(d-3)(d-4)kb}{r^2f}
-\frac{(d-1)(d-4)b}{ \ell^2f}
+\frac{(d-3)ba'}{ra}
\\
\nonumber
&&+\frac{(d-4)b'}{r}
-\frac{(d-4)kb'}{rf}
-\frac{(d-1)rb'}{\ell^2f}
+\frac{a'b'}{2a}
+\frac{b'^2}{b}+\frac{2q^2}{r^{2(d-3)}}\frac{(3d-8)b+rb'}{(d-2)af}~,
\end{eqnarray}
\begin{eqnarray}
\label{ep3} 
\nonumber
\frac{a'}{a}= 2\frac{ b\big[\ell^2(d-3)(d-4)(k-f)+(d-1)(d-2)r^2\big]
-(d-3)r\ell^2fb'}{r\ell^2f\big[rb'+2(d-3)b\big]}-\frac{4q^2}{r^{2d-7}}
\frac{b}{fa\big[rb'+2(d-3)b\big]}~.
\end{eqnarray}
For $k=0$, the EM equations admit the exact solution $a=r^2$, 
$f=1/b=-2m/r^{d-3}+r^2/\ell^2+2q^2/((d-2)(d-3)r^{2(d-3)})$, 
$V(r)=q/(r^{d-3}(3-d))+\Phi$,
which was recovered by our numerical 
procedure. This $k=0$ solution appears to be unique, corresponding to the 
known planar EM topological black hole.
 Therefore in the remainder of this section we will concentrate on 
the $k=\pm 1$  cases only.

\subsection{Asymptotics}
We consider solutions of the above equations whose boundary topology  
is the product of time and $S^{d-3}\times S^1$ or $H^{d-3}\times 
S^1$. As in the case of the neutral black string, we find that, 
for even $d$, the metric functions
admit at large $r$ a power series expansion of the form:
\begin{eqnarray} 
\nonumber
a(r)&=&\frac{r^2}{\ell^2}+\sum_{j=0}^{(d-4)/2}a_j(\frac{\ell}{r})^{2j}
+c_z(\frac{\ell}{r})^{d-3}+O(1/r^{d-2}),
\\
\label{even-inf}
b(r)&=&\frac{r^2}{\ell^2}+\sum_{j=0}^{(d-4)/2}a_j(\frac{\ell}{r})^{2j}
+c_t(\frac{\ell}{r})^{d-3}+O(1/r^{d-2}),
\\
\nonumber
f(r)&=&\frac{r^2}{\ell^2}+\sum_{j=0}^{(d-4)/2}f_j(\frac{\ell}{r})^{2j}
+(c_z+c_t)(\frac{\ell}{r})^{d-3}+O(1/r^{d-2}),
\end{eqnarray}   
where $a_j,~f_j$ are constants depending on the index
$k$ and the spacetime dimension only. Specifically, we find
\begin{eqnarray}
\label{inf2}  
a_0=(\frac{d-4}{d-3})k~,~~
a_1=\frac{(d-4)^2k^2}{(d-2)(d-3)^2(d-5)},~~
a_2=-\frac{(d-4)^3(3d^2-23d+26)k^3}{3(d-2)^2(d-3)^3(d-5)(d-7)}~,~~~
\end{eqnarray} 
\begin{eqnarray}
\label{inf3}  
f_0=\frac{k(d-1)(d-4)}{(d-2)(d-3)},~~
f_1=2a_1,~~
f_2=-\frac{2(d-4)^3(d^2-8d+11)k^3}{(d-2)^2(d-3)^3(d-5)(d-7)}~,
\end{eqnarray} 
their expression becoming more complicated for higher $j$, with no general
pattern becoming apparent.

The corresponding expansion for odd values of the spacetime 
dimension is given by\footnote{Note that in both even and odd dimensions one 
finds the asymptotic
expression of the
Riemann tensor
$R_{\mu \nu}^{~~\lambda \sigma}=-(\delta_\mu^\lambda \delta_\nu^\sigma
-\delta_\mu^\sigma \delta_\nu^\lambda)/\ell^2+O(1/r^{d-4})$.}
\begin{eqnarray}
\nonumber 
a(r)&=&\frac{r^2}{\ell^2}+\sum_{j=0}^{(d-5)/2}a_j(\frac{\ell}{r})^{2j}
+\zeta\log(\frac {r}{\ell}) (\frac{\ell}{r})^{d-3}
+c_z(\frac{\ell}{r})^{d-3}+O(\frac{\log r}{r^{d-1}}),
\\
\label{odd-inf}
b(r)&=&\frac{r^2}{\ell^2}+\sum_{j=0}^{(d-5)/2}a_j(\frac{\ell}{r})^{2j}
+\zeta\log (\frac {r}{\ell}) (\frac{\ell}{r})^{d-3}
+c_t(\frac{\ell}{r})^{d-3}+O(\frac{\log r}{r^{d-1}}),
\\
\nonumber
f(r)&=&\frac{r^2}{\ell^2}+\sum_{j=0}^{(d-5)/2}f_j(\frac{\ell}{r})^{2j}
+2\zeta\log (\frac {r}{\ell}) (\frac{\ell}{r})^{d-3}
+(c_z+c_t+c_0)(\frac{\ell}{r})^{d-3}+O(\frac{\log r}{r^{d-1}}),
\end{eqnarray}   
where we note $\zeta=a_{(d-3)/2}\sum_{k>0}(d-2k-1)\delta_{d,2k+1}$,
while
\begin{eqnarray}
\label{inf4}
c_0=0~~~{\rm for }~~d=5, ~~
c_0=\frac{9k^3\ell^4}{1600}~~~{\rm for }~~d=7, 
~~
c_0=-\frac{90625 k^4\ell^6}{21337344}~~~{\rm for }~~d=9. 
\end{eqnarray} 
For any value of $d$, terms of higher order in $ \ell/r$ depend on the two 
constants $c_t$ and $c_z$ and also on the charge parameter $q$. 
The  constants  $(c_t,~c_z)$ are found numerically starting from the following 
expansion of the solutions near the event horizon (taken at constant $r=r_h$) 
and integrating the EM equations towards infinity: 
\begin{eqnarray}
\nonumber  
a(r)&=&
a_h
+\frac{2a_hr_h\big[a_h(d-1)(d-2)r_h^{2d}-2q^2r_h^6\ell^2\big]}
{a_h(d-2)r_h^2d\big[(d-1)r_h^2+k(d-4)\ell^2\big]-2q^2\ell^2r_h^8}(r-r_h)
+\bar a_2(r-r_h)^2
+O(r-r_h)^3,
\\
\label{eh}
b(r)&=&b_1(r-r_h)
+\bar b_2(r-r_h)^2
+O(r-r_h)^3,
\\
\nonumber
f(r)&=&\frac{1}{r_h \ell^2}\bigg[(d-1)r_h^2+k(d-4)\ell^2-\frac{2q^2\ell^2}
{a_h(d-2)r_h^{2(d-4)}}\bigg](r-r_h)
+\bar f_2(r-r_h)^2
+O(r-r_h)^3,
\end{eqnarray}
in terms of two parameters $a_h$, $b_1$ (this expansion
corresponds to a nonextremal solution). The quantities $\bar a_2$, 
$\bar b_2$ and $\bar f_2$ can easily be found out by expanding Einstein's equations 
near horizon. We found that they are given by complicated expressions 
in terms of $a_h$, $b_1$ and $q$ and for simplicity we will not list them here. 
Let us also note that we considered the following expansion of the 
electromagnetic potential near horizon:
\beqs
\nonumber
V(r)&=&V_0+\frac{q\ell}{r_h^{d-3}}\sqrt{ \frac{b_1r_h}{a_h}}
\bigg({(d-1)r_h^2+k(d-4)\ell^2-\frac{2q^2\ell^2}
{a_h(d-2)r_h^{2(d-4)}}\bigg)^{-1/2}}(r-r_h)+O(r-r_h)^2,
\eeqs
where $V_0$ is a constant. 
Notice now that one can always set
$V_0=0$ such that $A_t(r_h)=0$. The physical significance of the 
quantity $\Phi$ in (\ref{Vr}) is then that it plays the role of the electrostatic 
potential difference between the infinity and  horizon.

 The condition for a regular event horizon is $f'(r_h)>0$, with $b'(r_h)>0$ 
and we find that $r_h$ must satisfy the equation:
 \beqs
(d-1)r_h^{2d-6}+ k(d-4)\ell^2r_h^{2d-8}\geq\frac{2q^2\ell^2}{a_h(d-2)}.
\label{cond}
\eeqs
For the uncharged string with $k=-1$,
this implies the existence of 
a minimal value of $r_h$, $i.e.$ for a given $\Lambda$,
$r_h>\ell\sqrt{(d-4)/(d-1)}.$
One can see that this is also a lower bound in the charged case as well,
 since in (\ref{cond}) one has $a_h>0$ and therefore the 
left hand side should be positive. If the equality in 
(\ref{cond}) is saturated the black hole horizon is 
degenerate and it corresponds to an extremal electrically 
charged black string. 
For the usual RN black hole in AdS backgrounds, 
a similar inequality imposes a bound on the mass of 
the black hole, which is again 
saturated for the extremal solution \cite{Chamblin:1999tk}.
Note also, using (\ref{Vr}) and (\ref{eh}), that no
reasonable charged solution can exist in the $r_h\to 0$ limit.

\subsection{Properties of the charged solutions}
The global charges of these solutions are computed by using the
counterterm formalism 
presented in Section 2.
The computation of the boundary stress-tensor
$T_{ab}$ is straightforward and we find the 
following expressions for mass and tension:\footnote{Note that 
 one can define a dimensionless relative tension
\cite{Harmark:2003eg}
$n={\mathcal T}L/M$, which is constant for $\Lambda=0$ uniform solutions. 
However, this quantity is configuration dependent for the 
 AdS black strings.} 
\begin{eqnarray}
\label{MT} 
M&=&M_0+M_c^{(k,d)}~,~~M_0=\frac{\ell^{d-4}}{16\pi G 
}\big[c_z-(d-2)c_t\big]L{\cal V}_{k,d-3}~,
\\
{\mathcal T}&=&{\mathcal T}_0+{\mathcal T}_c^{(k,d)}~,~~
{\mathcal T}_0=\frac{\ell^{d-4}}{16\pi G }\big[(d-2)c_z-c_t\big] 
{\cal V}_{k,d-3}~,
\end{eqnarray}  
where ${\cal V}_{k,d-3}$ is the total area of the angular sector. Here 
$M_c^{(k,d)}$ and ${\mathcal T}_c^{(k,d)}$ are  Casimir-like terms
which appear for an odd spacetime dimension only,
 \begin{eqnarray}
\label{MT-Cas} 
M_c^{(k,d)}=-L{\mathcal T}_c^{(k,d)} =\frac{\ell^{d-4}}{16\pi G 
}{\cal V}_{k,d-3}L\left(\frac{1}{12}\delta_{d,5}
-\frac{63}{1600}\delta_{d,7}+\dots\right)~.
\end{eqnarray}  
Using (\ref{Qc}) the electric charge is found to be:
\beqs
Q=-\frac{qL{\cal V}_{k,d-3}}{4\pi G_d}~.
\eeqs
The Hawking temperature $T_H=\kappa_H/2\pi$ as obtained from
the standard relation (\ref{kappa}) is: 
\begin{eqnarray}
T_H&=&
\frac{1}{4\pi}\sqrt{\frac{b_1}{r_h\ell^2}
\bigg((d-1)r_h^2+k(d-4)\ell^2-\frac{2q^2\ell^2}{a_h(d-2)r_h^{2(d-4)}}\bigg)}~.
\label{temp}
\end{eqnarray}
The area $A_H$ of the black string horizon is given by 
\begin{eqnarray}
\label{A} 
A_H=r_h^{d-3}{\cal V}_{k,d-3}L\sqrt{a_h}~.
\end{eqnarray}
As usual, one identifies the entropy of black string solutions
with one quarter of the even horizon area\footnote{
This relation has been derived in \cite{ Mann:2006yi}
by using Euclidean techniques. This is not  
possible for these electrically charged
solutions, which do not solve the equations of motion for an Euclidean signature. 
However, it can be deduced using the quasi-Euclidean approach.}, $S=A_H/4G_d$.

It is also possible to write a simple Smarr-type formula, relating quantities defined at 
infinity to quantities defined at the event horizon.
%
\begin{figure}[h!]
\parbox{\textwidth}
{\centerline{
\mbox{
\epsfysize=10.0cm
\includegraphics[width=92mm,angle=0,keepaspectratio]{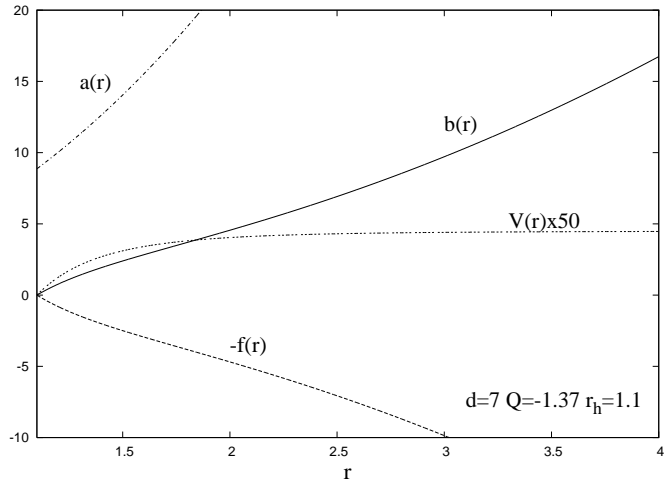} 
}}}
\caption{{\small The profiles of the metric functions $a(r)$, $b(r),~f(r) $ and 
 the electric gauge potential
 $V(r)$ are shown for a typical $d=7,~k=1$ charged black string solution.}}
\end{figure}
%
%
To this aim, we integrate the Killing 
identity $\nabla^\mu\nabla_\nu  \zeta_\mu=R_{\nu \mu}\zeta^\mu,$
for the Killing vector $\zeta^\mu=\delta^\mu_t$, together with the  Einstein 
equation $R_t^t={(R - 2\Lambda+F^2)/2}$. Next step is to evaluate the
tree level action (\ref{action}) by isolating the 
bulk action
contribution at infinity and at $r=r_h$. The divergent 
contributions given by the surface integral term at infinity are also 
canceled by $I_{\rm{surface}}+I_{ct}$.  
The same approach applied to the Killing vector $\zeta^\mu=\delta^\mu_z$ yields 
the result:
\begin{eqnarray}
\label{itot2}
I =-\beta {\mathcal T}L.
\end{eqnarray}
which together with  
 \begin{equation}
T_H (S+I)=  M-\Phi Q .  
\label{GibbsDuhem}
\end{equation}%
lead to the 
Smarr-type formula 
\begin{eqnarray}
\label{smarrform} 
M+{\mathcal T}L-\Phi Q=T_HS~.
\end{eqnarray} 

\subsection{Numerical results}
Although an analytic or approximate solution 
of the equations (\ref{ep1}))
 appears to be intractable, here we 
present arguments for the existence of nontrivial solutions,
which smoothly interpolate between the 
asymptotic expansions (\ref{eh}) and (\ref{odd-inf}) or 
(\ref{even-inf}). 
The numerical techniques we used to find charged 
black string solutions are similar to  
those employed in the vacuum case.

 Starting 
from the even horizon expansion (\ref{eh}) (with $V(r_h)=0$)
and using a standard ordinary differential  
equation solver, we integrated the EM equations 
adjusting  for shooting parameters and 
integrating towards  $r\to\infty$. The 
integration stops when the asymptotic limit 
(\ref{even-inf}), (\ref{odd-inf}) is reached
with a reasonable accuracy. Given $(k,~d,~\Lambda,~q,~r_h)$, 
solutions with the right asymptotics exist 
for  one set of the shooting parameters  $(a_h,~b_1)$ only.

The results we present here are found for $\ell=1$. 
However, the solutions for any other values of the cosmological 
constant are easily found by using a suitable rescaling of the 
$\ell=1$ 
configurations. 
\begin{figure}[h!]
\parbox{\textwidth}
{\centerline{
\mbox{
\epsfysize=10.0cm
\includegraphics[width=82mm,angle=0,keepaspectratio]{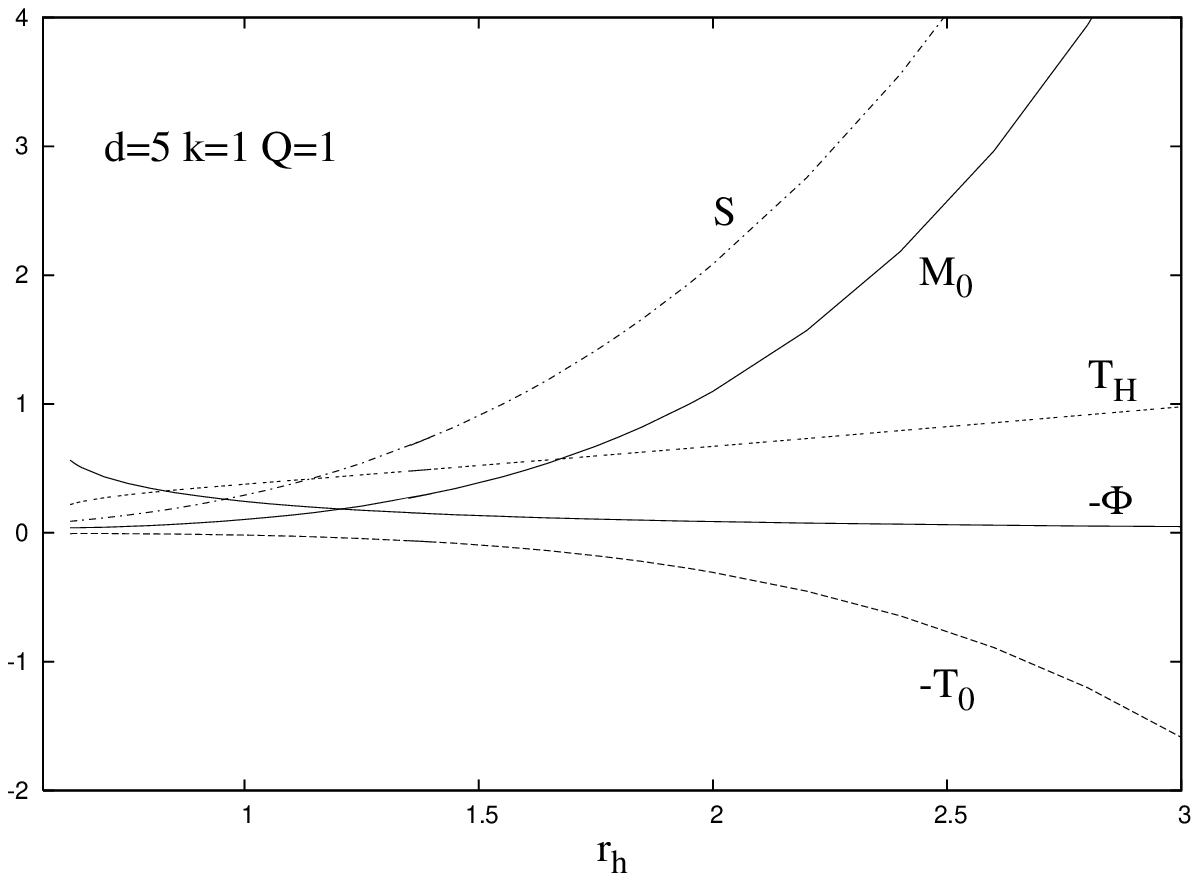}
\includegraphics[width=82mm,angle=0,keepaspectratio]{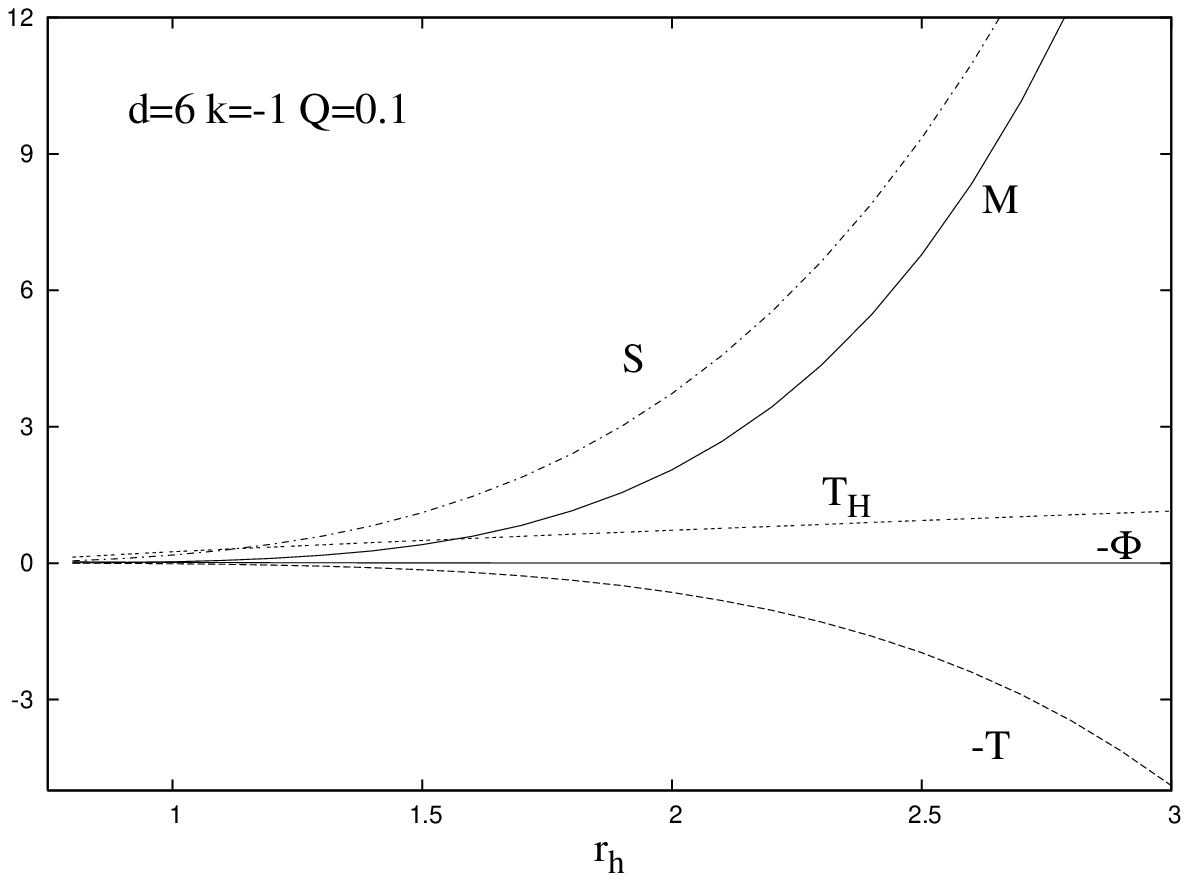}
}}}
\caption{{\small The mass-parameter $M$, 
 the tension ${\mathcal T}$,  the Hawking 
 temperature $T_H$ and the entropy $S$  as well as the 
 value of the electric potential at infinity $\Phi$ are 
 represented as functions of the event horizon radius 
 for $d=5, k=1$, respectively $d=6,~k=-1$ black string 
 solutions with fixed values of the electric charge.  
 Note that here and in Figure 3 we have subtracted the Casimir terms  for $d=5$
configurations.}}
\end{figure}
%
Indeed, to understand the dependence 
of the solutions on the cosmological constant, we note that the 
EM equations (\ref{ep2})) 
are left invariant by the transformation: 
\begin{eqnarray}
\label{transf1} 
r \to \bar{r}= \lambda r,~~\ell \to 
\bar{\ell}= \lambda \ell,~~q \to \bar{q}= \lambda^{d-4} q,
\end{eqnarray}
Therefore, starting from a solution corresponding to 
$\ell=1$ one may generate in this way a 
family of $\ell\neq 1$ solutions, which, 
similarly to the vacuum case, are 
termed ``copies of solutions`` \cite{Harmark:2003eg}. 
The new solutions have the same length in the extra-dimension. 
Their relevant properties, expressed in terms of 
the corresponding properties of the initial solution, are as follows:
\begin{eqnarray}
\label{transf2} 
 \bar{r}_h=\lambda r_h,~\bar{\Lambda}=\Lambda/\lambda^2 ,~~
 \bar{T}_H=T_H/\lambda ,~~
 \bar{M}=\lambda^{d-4} M ,~~ \bar{Q}=\lambda^{d-4} Q ,~~{\rm and}~~
 \bar{{\mathcal T}}=\lambda^{d-4} {\mathcal T}.~~~{~~}
\end{eqnarray}
We have found numerically charged static black strings solutions 
with AdS asymptotics in all dimensions between five and ten. 
They are likely to exist for any $d\geq 5$. 
For all the solutions 
we studied, the metric functions $a(r)$, $b(r)$ , $f(r)$ 
and the electric gauge potential $V(r)$ interpolate
 monotonically between the corresponding values at $r=r_h$ and the 
asymptotic values at infinity, without presenting any local extrema. 
As a typical example, in Figure $1$ the metric functions $a(r)$, 
$b(r)$ and $f(r)$ as well as the electric gauge potential $V(r)$ 
are shown as functions of the radial coordinate $r$ for a $d=7,~k=1$  
solution with  $r_h=1.1,~Q=1.37$. One can see that the term $r^2/\ell^2$ 
starts dominating the profile of these functions very rapidly, 
which implies a small difference between the metric functions for 
large enough $r$, while the gauge potential $V(r)$ approaches very fast a constant value.
 
The dependence of various physical parameters on the event horizon 
radius is presented in Figure 2 for $d=5,~k=1$ and $d=6,~k=-1$ solutions 
with fixed values of $Q$. These plots retain the basic features 
of the solutions we found in other dimensions (note that in 
this 
paper we set  $L={\cal V}_{k,d-3}/G_d=1$ in the numerical values for the mass, 
tension, angular momentum and entropy and $L{\cal V}_{k,d-3}/4\pi G_d=1$
for charge;
also we subtracted the Casimir energy and tension in odd dimensions). 
For a given value of the electric charge, the mass,  
tension and entropy of charged solutions increase 
monotonically with $r_h$.
Alternatively, one may keep fixed the value $\Phi$ of 
the electric potential at infinity.
%
%
\begin{figure}[h!]
\parbox{\textwidth}
{\centerline{
\mbox{
\epsfysize=10.0cm
\includegraphics[width=82mm,angle=0,keepaspectratio]{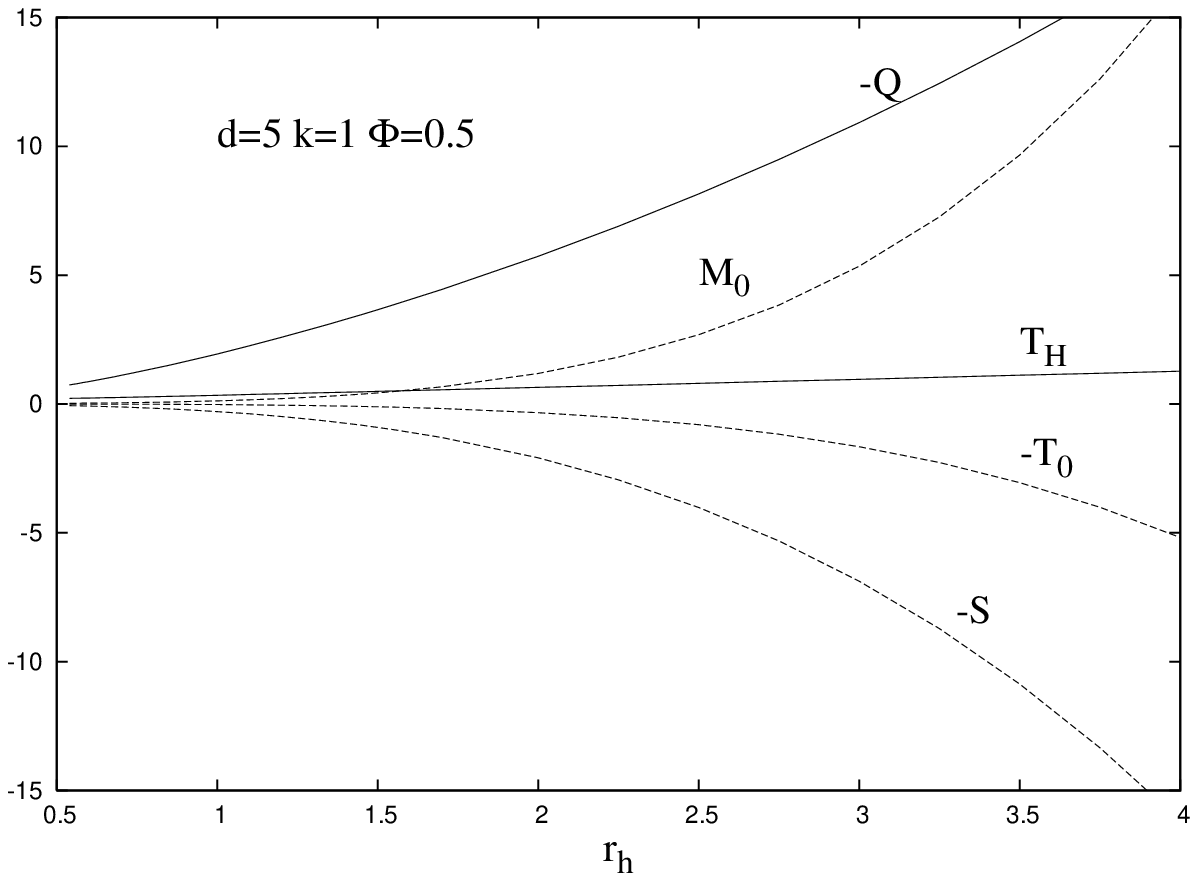}
\includegraphics[width=82mm,angle=0,keepaspectratio]{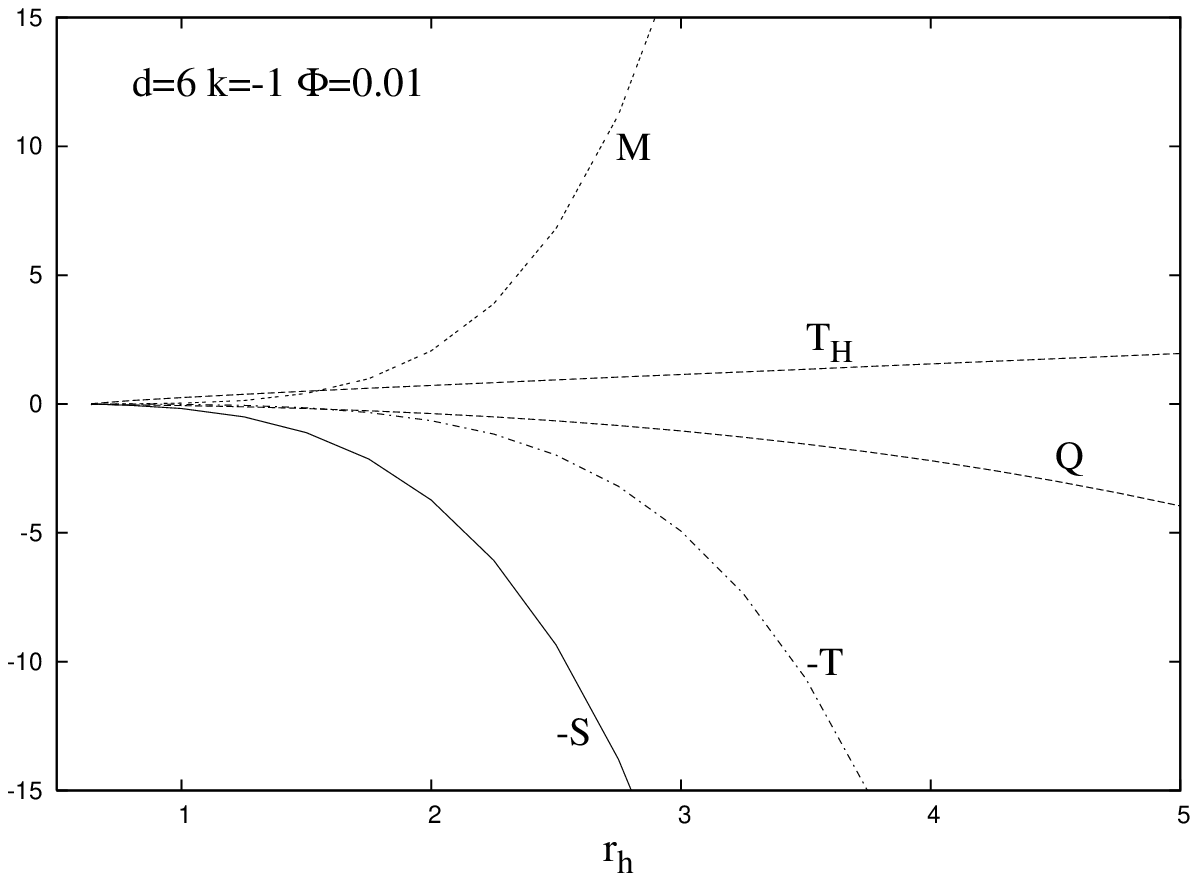}
}}}
\caption{{\small The mass-parameter $M$, 
 the tension ${\mathcal T}$, the Hawking 
 temperature $T_H$, the entropy $S$  as well as the 
 electric charge $Q$ are 
 represented as functions of the event horizon radius 
 for $d=5, k=1$, respectively $d=6,~k=-1$  black string 
 solutions with fixed values of the electric potential at infinity.}}
\end{figure}
%
%
The corresponding picture in this case is shown in Figure 3
for a set of $d=5 $ solutions with an $S^2\times S^1$ topology of the 
event horizon.

As expected, for any value of $Q\neq 0$, we notice the presence of 
a minimal value $r_h^{(m)}$ of the event horizon radius.
As $r_h \to r_h^{(m)}$ a critical solution is approached
and the numerical solver fails to converge.
The study of this critical 
solution seems to require a different
parametrization of the metric ansatz and is beyond the purposes of this paper. 
 Also, for both $k=\pm 1$ cases, we did not notice 
the existence of a maximal allowed value of the event horizon radius.

\section{Rotating neutral solution with equal magnitude angular momenta} 
\subsection{The  equations and boundary conditions}
 
To obtain stationary black string solutions,
representing rotating generalizations of the $k=1$ static
 solutions discussed in \cite{Mann:2006yi},
we consider even dimensional space-times with  $N=(d-2)/2$ commuting Killing vectors 
$\eta_k=\partial_{\varphi_k}$.

We employ a parametrization for the metric,
well suited for numerical work,
corresponding to a generalization of the Ansatz  
previously used for static  black strings\footnote{
A closely related ansatz has been used to find higher dimensional
generalizations of the Kerr-Newman black hole solutions
with equal magnitude angular momenta in asymptotically flat \cite{Kunz:2006eh}
and AdS \cite{Kunz:2007jq} spacetimes.}
\begin{eqnarray}
& ds^2 = -b(r)dt^2 +a(r)dz^2+  \frac{ dr^2}{f(r)} + 
g(r)\sum_{i=1}^{N-1}
  \left(\prod_{j=0}^{i-1} \cos^2\theta_j \right) d\theta_i^2  
\nonumber
 \\  
&+h(r) \sum_{k=1}^N \left( \prod_{l=0}^{k-1} \cos^2 \theta_l
  \right) \sin^2\theta_k \left( d\vphi_k - w(r)
  dt\right)^2 
  \label{metric-rot}
 \\ 
& +(g(r)-h(r)) \left\{ \sum_{k=1}^N \left( \prod_{l=0}^{k-1} \cos^2
  \theta_l \right) \sin^2\theta_k  d\vphi_k^2 \right. 
  -\left. \left[\sum_{k=1}^N \left( \prod_{l=0}^{k-1} \cos^2
  \theta_l \right) \sin^2\theta_k   d\vphi_k\right]^2 \right\},
\nonumber
\end{eqnarray}
where $\theta_0 \equiv 0$, $\theta_i \in [0,\pi/2]$ 
for $i=1,\dots , N-1$, $\theta_N \equiv \pi/2$, 
$\vphi_k \in [0,2\pi]$ for $k=1,\dots , N$.
Note that the static black strings in even dimensions 
discussed in \cite{Mann:2006yi} are recovered for $w(r)=0$, $h(r)=g(r)=r^2$.

A convenient metric gauge choice in the numerical procedure is $h(r)=r^2$. 
Thus we find the following field equations: 
\begin{eqnarray}
\label{eq1} 
\nonumber
&\frac{a'}{a}=-\bigg[2\ell^2 
fg\left(rgb'+b(2g+(d-4)rg')\right)\bigg]^{-1}
\bigg [ b\bigg(-4(d-1)(d-2)rg^2-4(d-4)\ell^2g\big((d-2)r-fg'\big)~~{~~~~}
\\
\nonumber
&+(d-4)r\ell^2(4r^2+(d-5)fg'^2)\bigg)
+2 \ell^2 fg((d-4)rb'g'+g(2b'+r^3w'^2))\bigg]~,~~~{~~~~~~}
\end{eqnarray}
\begin{eqnarray}
\nonumber
&f'=\frac{1}{d-2}\bigg(
\frac{(d-4)(2d-3)r^3}{g^2}
-\frac{(d-4)( d-2)r }{g }
+\frac{(d-1)(d-2)r }{\ell^2}
+(\frac{ a'}{a}+\frac{b'}{b})((d-4)\frac{rg'}{2g}
\\
\label{eq2} 
&-d+3)f+\frac{ rfa'b'}{2ab}
-\frac{(d-3)(d-4)fg'}{g}
+\frac{(5-2d)r^3f}{2b}w'^2
+\frac{(d-5)(d-4)rf}{4g^2}g'^2 
 \bigg)~,~~~{~~~~~~}
\end{eqnarray}
\begin{eqnarray}
\nonumber
&b''=\frac{1}{d-2}\bigg(
\frac{(d-4)(d-5)b}{4g^2}g'^2
+\frac{(2d-3)r^2}{2}w'^2
-\frac{(d-3)(d-4) }{2g}b'g'
+\frac{(d-4) b}{2ag}a'g'
-\frac{(d-2) }{2f}b'f'~~~~~~~~~{~~~~~~~}
\\
\label{eq3} 
\nonumber
&+\frac{(d-2)b'^2}{2b}
-\frac{(d-3)}{2a }a'b'
+\frac{ba'}{ra} 
-\frac{(d-3)b' }{r} 
+\frac{(d-4) bg'}{rg} 
+\frac{(d-4)r^2b}{fg^2}-\frac{(d-2)(d-4)b}{fg} 
+\frac{(d-1)(d-2)b}{\ell^2f} 
\bigg)~,~~~~~~~~{~~~~~~~~~~}
\end{eqnarray}
\begin{eqnarray}
\nonumber
& g''=\frac{1}{d-2}\bigg(
\frac{r^2g}{2b}w'^2-\frac{d^2-7d+4}{4g}g'^2
-(d-2)\frac{f'g'}{2f}
+(\frac{a'}{a}+\frac{b'}{b})(\frac{g}{r}-g')
+\frac{ga'b'}{2ab}
-\frac{2g'}{r}
\\
\nonumber
&+\frac{(4-3d)r^2}{fg}
+\frac{d(d-2)}{f}
+\frac{(d-1)(d-2)g}{\ell^2 f}
\bigg),
\end{eqnarray}
\begin{eqnarray}
\label{eqw}
\nonumber
& (g^{\frac{d-4}{2}} r^3\sqrt{\frac{af}{ b}}w')'=0~.
\end{eqnarray}
The last equation in the relations above implies the existence of the 
first integral
\begin{eqnarray}
\label{fiw}
w'=\alpha g^{-\frac{d-4}{2 }}\frac{1}{r^3} \sqrt{ \frac{b}{af }},
\end{eqnarray}
where $\alpha$ is a constant fixing the total angular momentum $J$ of 
the solutions.
\subsection{Asymptotic expansion and physical quantities}
We are interested in black string solutions, with an 
horizon located at a constant value of
the radial coordinate
$r=r_h$. Restricting again to
nonextremal solutions, the following expansion holds near $r=r_h$:
\begin{eqnarray}
\nonumber
&&f(r)=\bar f_1(r-r_h)+\bar f_2(r-r_h)^2+O(r-r_h)^3,~~g(r)=g_h+g_1(r-r_h)+O(r-r_h)^2,
\\
\label{eh2}
&&a(r)=a_h+ \bar a_1(r-r_h)+O(r-r_h)^2,~~b(r)=  b_1(r-r_h)+\bar b_2(r-r_h)^2+O(r-r_h)^3,
\\
\nonumber
&&w(r)=w_h+w_1(r-r_h)+O(r-r_h)^2,
\end{eqnarray}
in terms of four essential parameters $a_h$, $\bar b_1$,~$w_h$ and $g_h$. 
The remaining quantities $\bar a_1$, $\bar b_2$,   $w_1$ and $\bar f_2$ 
that appear in the above near-horizon expansion can be found out 
by solving Einstein's equations near horizon and are given by 
complicated expressions in terms of $a_h$, $\bar b_1$,~$w_h$ and $g_h$. 
The metric functions have the following asymptotic behaviour in 
terms of four arbitrary constants $c_t,~c_z,~c_g$ and $c_\varphi$:
\begin{eqnarray} 
\nonumber
&&a(r)=\frac{r^2}{\ell^2}+\sum_{j=0}^{(d-4)/2}a_j(\frac{\ell}{r})^{2j}
+c_z(\frac{\ell}{r})^{d-3}+O(1/r^{d-2}),
\\
\nonumber
&&b(r)=\frac{r^2}{\ell^2}+\sum_{j=0}^{(d-4)/2}a_j(\frac{\ell}{r})^{2j}
+c_t(\frac{\ell}{r})^{d-3}+O(1/r^{d-2}),
\\
\label{even-inf-rot}
&&f(r)=\frac{r^2}{\ell^2}+\sum_{j=0}^{(d-4)/2}f_j(\frac{\ell}{r})^{2j}
+(c_z+c_t+(d-4)c_g)(\frac{\ell}{r})^{d-3}+O(1/r^{d-2}),
\\
\nonumber
&&g(r)=  r^2(1+c_g (\frac{\ell}{r})^{d-1})+O(1/r^{d-1 }),
~~
w(r)=   c_\varphi (\frac{\ell}{r})^{d-1}+O(1/r^{d }),
\end{eqnarray}   
where $a_j,~f_j$ are constants still given by (\ref{inf2}), (\ref{inf3}) (with $k=1$).
For any even value of $d$, terms of higher order in $ {\ell}/{r}$ 
depend on the three
constants $c_t,~c_z$ and $c_g$ and also on the  parameter $c_{\varphi}$. 
The last relation in (\ref{fiw}) together with 
the asymptotic behaviour (\ref{even-inf-rot}) implies:
\begin{eqnarray}
\label{r3} 
\alpha=-c_{\varphi}(d-5)\ell^{d-4}.
\end{eqnarray}
Also, one can easily see that no
reasonable rotating solution  may exist in the $r_h\to 0$ limit.

As in the $(3+1)$-dimensional Kerr black hole case, the rotating black 
strings have an ergosurface inside of which the observers 
cannot remain stationary, and will move in the direction of rotation. 
The ergoregion is the region bounded by the event horizon, located 
at $r=r_h$ and the stationary limit surface, 
or the ergosurface, 
which is determined by the condition that on it the Killing 
vector $\partial/\partial t$ becomes null, $i.e.$ the surface given by $g_{tt}=0$, where:
\begin{eqnarray} 
\label{er}  
-g_{tt}&=&b(r)- r^2 w(r)^2 \sum_{k=1}^N \left( \prod_{l=0}^{k-1} \cos^2 \theta_l
  \right) \sin^2\theta_k=b(r)- r^2 w(r)^2,
\end{eqnarray} 
where in the last equality we have used the fact
 that in our ansatz $\theta_N=\pi/2$.
 To find out whether the ergosurface intersects the horizon, 
recall that from the near-horizon expansion 
(\ref{eh2}) we have on the horizon $b(r_h)=0$, 
while $w(r_h)=w_h\neq 0$ and, therefore, there 
are no intersection points with the horizon.

The Killing vector  $\chi=\partial/\partial_t+
\sum_k\Omega_k \partial/\partial \varphi_k $ is 
orthogonal to and null on the horizon. For the solutions 
within the ansatz (\ref{metric-rot}), the 
event horizon's angular velocities are 
all equal, $\Omega_k=\Omega_H=w(r)|_{r=r_h}$.
The Hawking temperature $T_H=\kappa_H/2\pi$ as  obtained from  (\ref{kappa}) is:
\begin{eqnarray} 
\label{Temp-rot} 
  T_H=\frac{\sqrt{b'(r_h)f'(r_h)}}{4\pi}.
\end{eqnarray} 
The area $A_H$ of the rotating black string horizon is given by: 
\begin{eqnarray}
\label{A2} 
A_H= r_h g_h^{(d-4)/2}{\cal V}_{1,d-3}L\sqrt{a_h}.
\end{eqnarray} 
%
\begin{figure}[h!]
\parbox{\textwidth}
{\centerline{
\mbox{
\epsfysize=10.0cm
\includegraphics[width=92mm,angle=0,keepaspectratio]{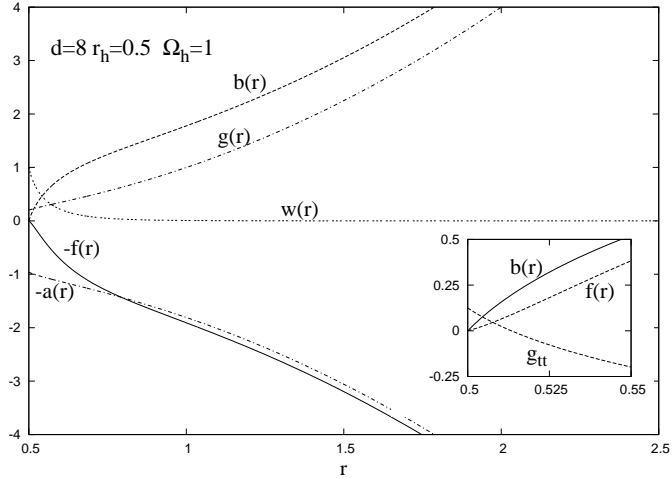} 
}}}
\caption{{\small The profiles of the metric 
functions $a(r)$, $b(r),~f(r),~g(r)$ and 
$w(r)$ are shown for a typical $d=8$ rotating black 
string solution with $r_h=0.5$, 
$\Omega_H=1$.}}
\end{figure}
%
Similarly to the static case, the conserved 
charged of the black strings are obtained by 
using the counterterms method in conjunction 
with the quasilocal formalism. Using the 
counterterms presented in Section $2$, the 
computation of 
$T_{ab}$ is straightforward and we find the 
following expressions for mass, tension and 
angular momentum\footnote{Note that these quantities are 
evaluated in a frame which is nonrotating
at infinity.}:  
\begin{eqnarray}
\label{MT-rot} 
&&M= \frac{\ell^{d-4}}{16\pi G_d 
}\big[c_z-(d-4)c_t+(d-4)c_g \big]L{\cal V}_{d-3}~,
\\
\nonumber
&&{\mathcal T}= \frac{\ell^{d-4}}{16\pi G_d }\big[(d-4)c_z-c_t-(d-2)c_g 
\big]
{\cal V}_{ {d-3}}~,~~~J_{(k)}=J=\frac{\ell^{d-4}}{16\pi G_d
}\frac{(d-1)c_\varphi}{d}{\cal V}_{d-3}.
\end{eqnarray} 
A computation similar to that presented in Section $3.3$ 
leads to the following
Smarr-type relation\footnote{The coefficient $(d-2)/2$ 
in the angular momentum contribution comes from
 $\Omega_1J_1+\Omega_2J_2+...+\Omega_NJ_N$, 
where $N=(d-2)/2$, with all these factors being equal in our case.}: 
\begin{eqnarray}
\label{smarrform-rot} 
M+{\mathcal T}L-\frac{(d-2)}{2}\Omega_HJ=T_HS,
\end{eqnarray} 
where as usual the entropy is $S=A_H/4G_d$.

\subsection{Numerical results}
The equations (\ref{eq1})-(\ref{eqw}) have been solved for several even values of $d$, 
special attention being paid to the case $d=6$. 
This was done since the decay of the relevant parts 
of the metric functions is lower in the six dimensional case
and this leads to better numerical accuracy; also, 
the properties of the $d=6$ case appear to be generic.
\begin{figure}[h!]
\parbox{\textwidth}
{\centerline{
\mbox{
\epsfysize=15.0cm
\includegraphics[width=82mm,angle=0,keepaspectratio]{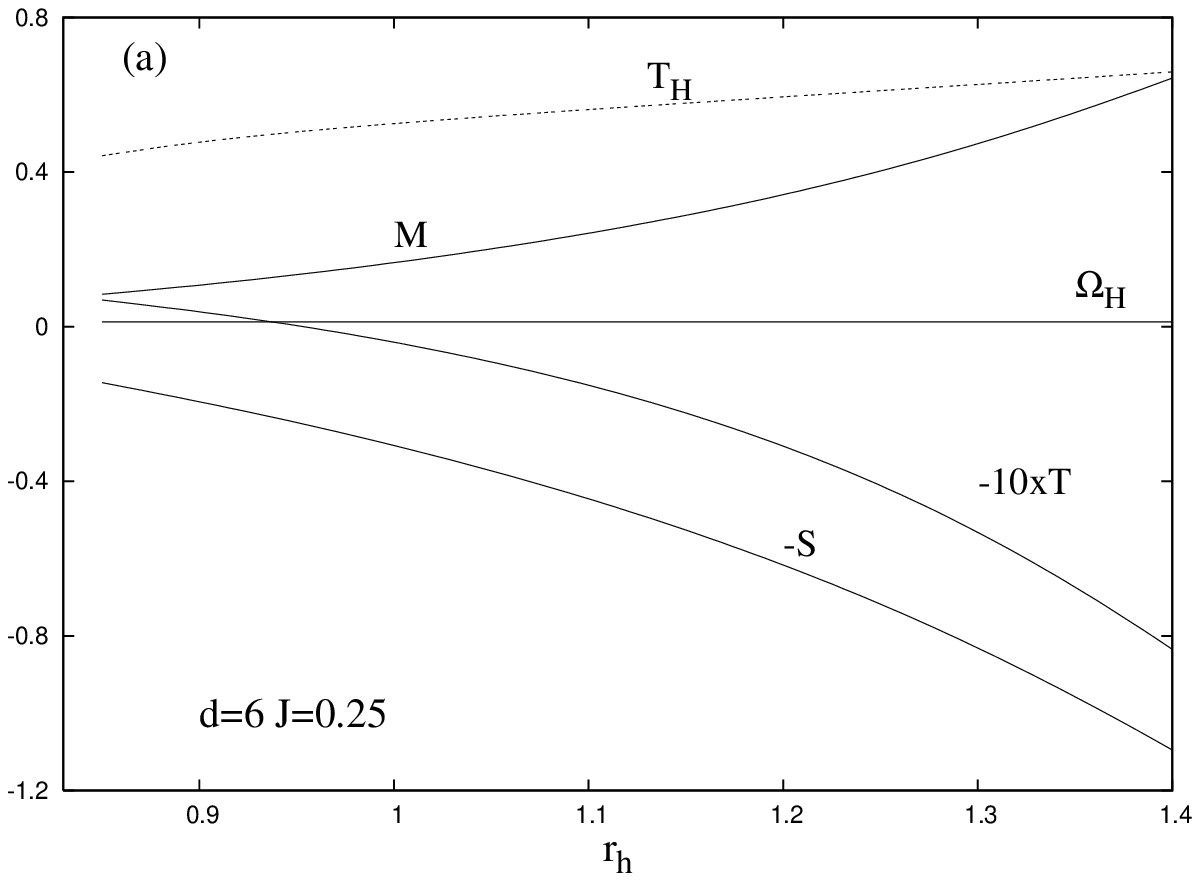}
\includegraphics[width=82mm,angle=0,keepaspectratio]{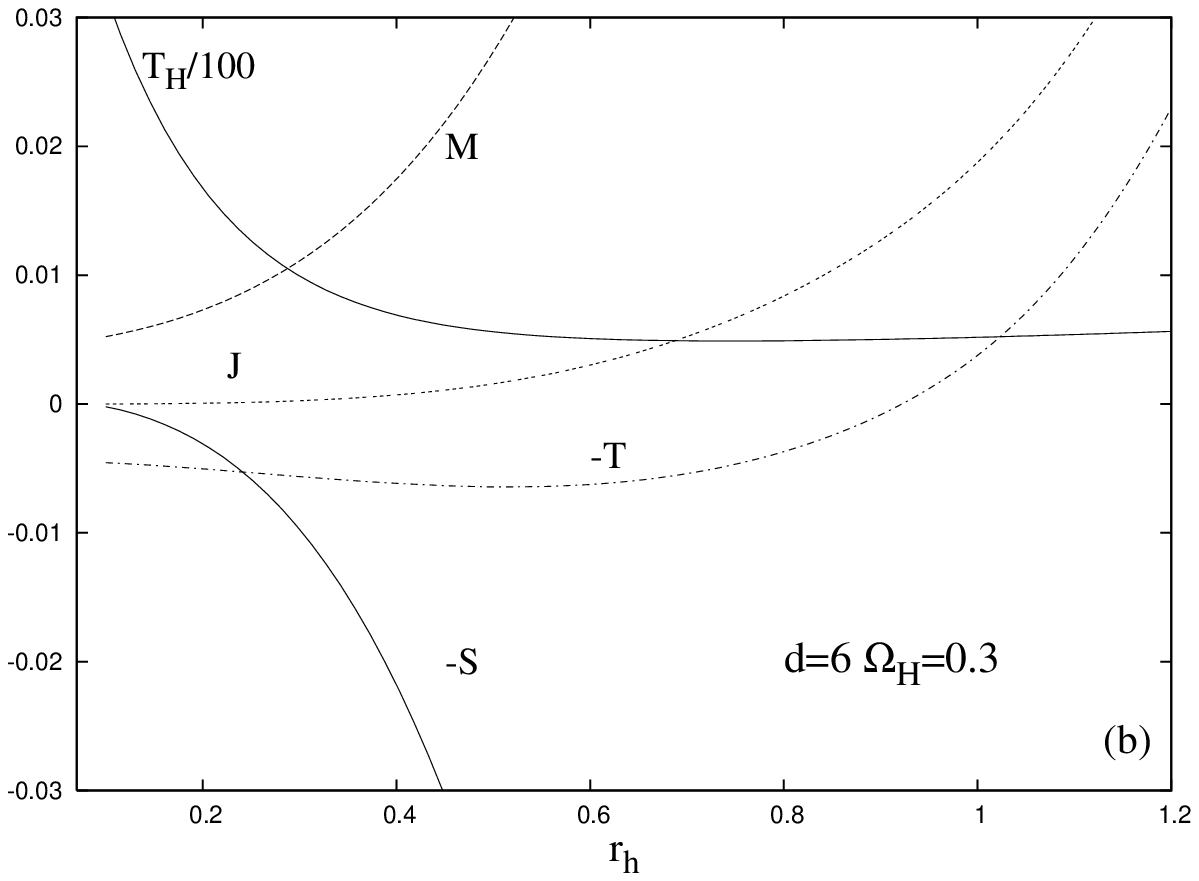}
}}}
\caption{{\small The mass-parameter $M$, the tension ${\mathcal T}$, 
the Hawking temperature $T_H$, the entropy $S$ as well as the angular momentum
 $J$ are represented as functions of the event horizon radius $r_h$ for $d=6$ 
rotating black string solutions with fixed angular momentum 
$J$ (Figure 5a) and fixed event horizon velocity $\Omega_H$ (Figure 5b).}}
\end{figure}

The numerical methods employed here are different from 
those used to describe the static charged black strings, 
and we used the differential equation solver $COLSYS$, 
which involves a Newton-Raphson method \cite{COLSYS}. 
Rotating solutions are found by starting with a static 
vacuum configuration with a given $r_h$ and increasing 
the value of $\Omega_H$ or $J$. The Einstein equations have been 
solved for a range of $r_h$ imposing the appropriate boundary 
conditions at the horizon and at infinity.

The set of boundary 
conditions is completely specified 
by fixing by hand either 
$\Omega_H$ or the angular momentum $J$ through fixing the 
quantity $c_{\varphi}$. We used both possibilities that, 
apart from leading to different physical issues for the 
solution, provide useful consistency cross-checks of the numerical method.

The complete classification of the solutions in the space 
of parameters is a considerable task that is 
not aimed in this paper. Instead, by taking $\ell=1$ 
we analyzed in detail a few particular classes of solutions, 
which hopefully would reflect all relevant properties of the general pattern.
Indeed, as in the charged case discussed in the previous section, 
the solutions for any other value of the cosmological constant are 
easily found by using a suitable rescaling of the $\ell=1$ configurations. 
To this end, and to understand the dependence of the solutions on 
the cosmological constant, we note that the Einstein equations 
in the rotating case (\ref{eq1})-(\ref{fiw}) are left invariant 
by the following transformation\footnote{A quick look at the 
field equations (\ref{eq1})-(\ref{fiw}), respectively at the 
asymptotic expansions of the metric functions, reveals that under 
this scaling symmetry $a, b, f$ remain invariant while $g\ra \lambda^2 g$ 
and $w\ra \lambda^{-1}w$. Notice that according to (\ref{r3}) 
we have $c_{\varphi}\ra \lambda^{-1}c_{\varphi}$.}: 
\begin{eqnarray}
\label{transf3a} 
r \to \bar{r}= \lambda r,~~\ell \to \bar{\ell}= \lambda \ell,
~~\alpha \to \bar{\alpha}= \lambda^{d-3} \alpha.
\end{eqnarray}
Therefore, starting from a solution corresponding to 
$\ell=1$ one may generate in this way a family of  
``copies of solutions" with $\ell\neq 1$, with the same length $L$ for the 
$z-$direction. 
Their relevant properties, expressed in terms of the 
corresponding properties of the initial solution, are as follows:
\begin{eqnarray}
\nonumber
\label{transf3b} 
 \bar{r}_h=\lambda r_h,~\bar{\Lambda}=\Lambda/\lambda^2 ,~
 \bar{T}_H=T_H/\lambda ,~\bar{\Omega}_H=\Omega_H/\lambda,~
\bar{M}=\lambda^{d-4} M ,~ \bar{J}=\lambda^{d-3} J ,~{\rm and}~
 \bar{{\mathcal T}}=\lambda^{d-4} {\mathcal T}.
\end{eqnarray}
We first discuss the general features of solutions for a
 fixed angular velocity of the event horizon. In this case, 
we were able to construct branches of solutions for increasing
 $r_h $ for several values of $\Omega_H$.
 These branches can alternatively be constructed by 
keeping $r_h$ fixed and increasing the parameter
$\Omega_H$. In this approach, we observe that the 
parameter $\vert w'(r_h) \vert$ becomes very
large (likely infinite) for some maximal value of $\Omega_H$. 
The precise evaluation of this value
is not easy because the numerical
analysis becomes very involved in this region; 
for example we obtain solutions up to  $\Omega_H \leq 1.22$ 
in the case $r_h=0.5$.
The maximal value of $\Omega_H$  decreases slowly while $r_h$ increases.

Again, the metric functions are monotonical functions of $r$, 
approaching fast the asymptotics (see Figure 4 for a plot of a 
typical $d=8$ rotating configuration).
Examining the component $g_{tt}$
of the metric (see (\ref{er})), we can determine the radius, say $r_e$, for
which $g_{tt}(r_e)=0$, determining the ergosurface.  
From the bounday conditions and (\ref{er})
we can see indeed that  $g_{tt}$ 
takes positive values for a some region $r_h<r<r_e$, as 
one can see in the inlet in Figure 4. 
 For a given $r_h$, 
the value of $r_e$ increases slightly  
with $\Omega_H$. For example, in the case $d=6$, we find: 
\begin{eqnarray} 
&r_e \simeq  0.508 \ , \ 0.525 \ , \  0.527\ \ &{\rm for} 
\ \ \Omega_H = 0.5 \ , 1.0 , \ 1.18 \ \ {\rm and } \ \  r_h = 0.5~,\ \ 
\nonumber \\
&r_e \simeq 1.028 \ , \ 1.037 \ , \  1.042\ \ &{\rm for} 
 \ \ \Omega_H = 0.5 \ , 0.7 , \ 0.85 \ \ {\rm and } \ \  r_h = 1.0~.
\nonumber
\end{eqnarray}
In the limit $r_h \to 0$, 
the rotating solution becomes singular, for any nonzero value of $\Omega_H$. 
This is indicated by the fact that some of the parameters, namely 
$b'(r_h),w'(r_h)$, become infinite in this limit while $J$ 
tends to zero. 

Studying the solutions in the plane $(r_h,J)$ with $J$ fixed 
reveals that, for $J >0$, black string solutions exist only for 
an  horizon radius larger than a minimal value, i.e. $r_h > r_{h}^{(m)}$
(depending on $J$). This observation is in full agreement with the property 
discussed 
above, which states that  no regular solution exist for $w \neq 0$.
For instance, for $J=0.025$ and $J=0.25$ we find  $r_{h}^{(m)} \approx 0.48$ and 
$r_{h}^{(m)} \approx 0.82$, respectively. 

Again, several quantities, namely $b(r_h), w(r_h), -w'(r_h)$, 
become very large in the limit $r_h \to r_{h}^{(m)}$.

 In Figure 5 we present the dependence of various physical parameters on the event horizon 
radius for $d=6$  solutions with a fixed value  of either $J$ or $\Omega_H$. 
These plots retain the basic features of the solutions we found in other dimensions.

 \section{Thermodynamics of the $k=1$ solutions}
 
As we have mentioned in Section $2$, the thermodynamics 
of the black objects is usually studied on the Euclidean section.
 However, for the solutions discussed in this paper it is not possible 
to find directly real solutions on the 
Euclidean section by Wick rotating $t\ra i\tau$ the Lorentzian 
configurations.
%
\begin{figure}[h!]
\parbox{\textwidth}
{\centerline{
\mbox{
\epsfysize=10.0cm
\includegraphics[width=82mm,angle=0,keepaspectratio]{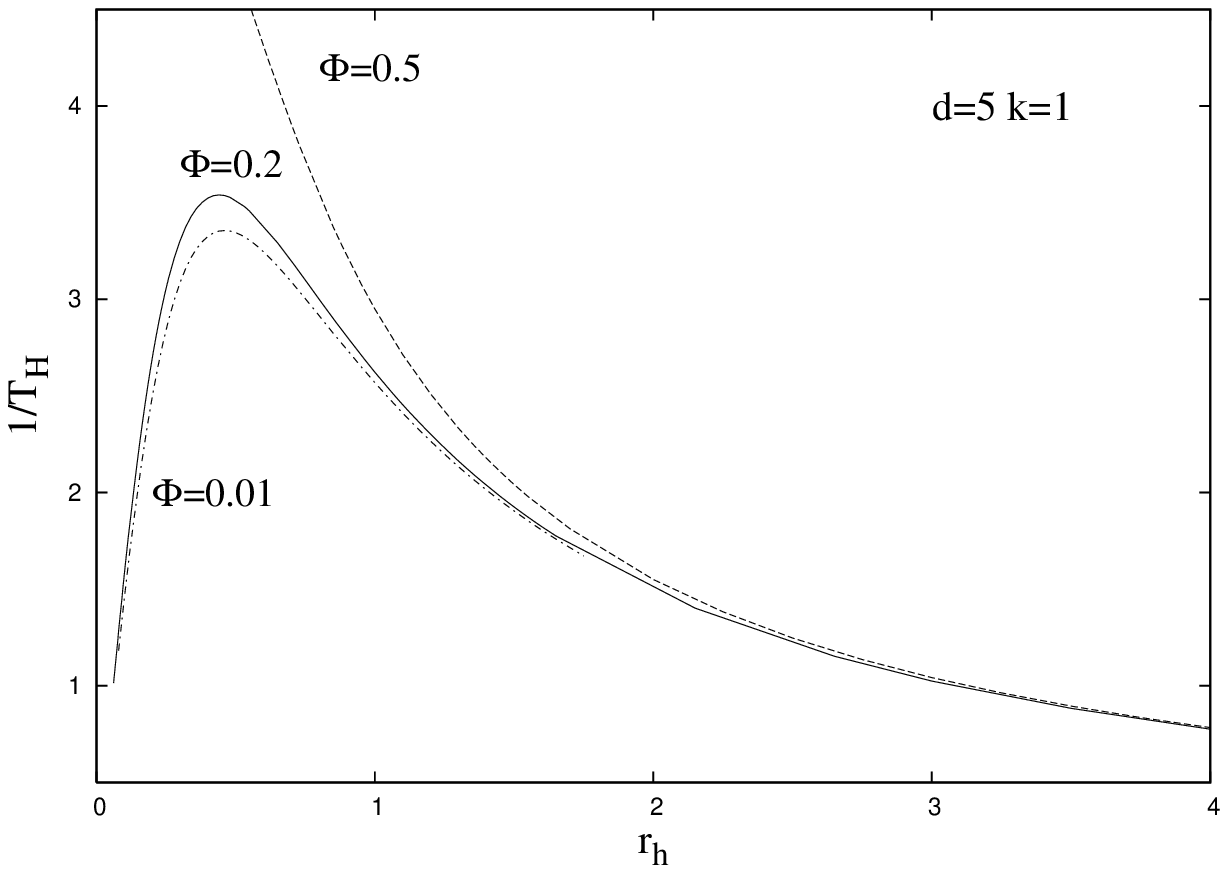}
\includegraphics[width=82mm,angle=0,keepaspectratio]{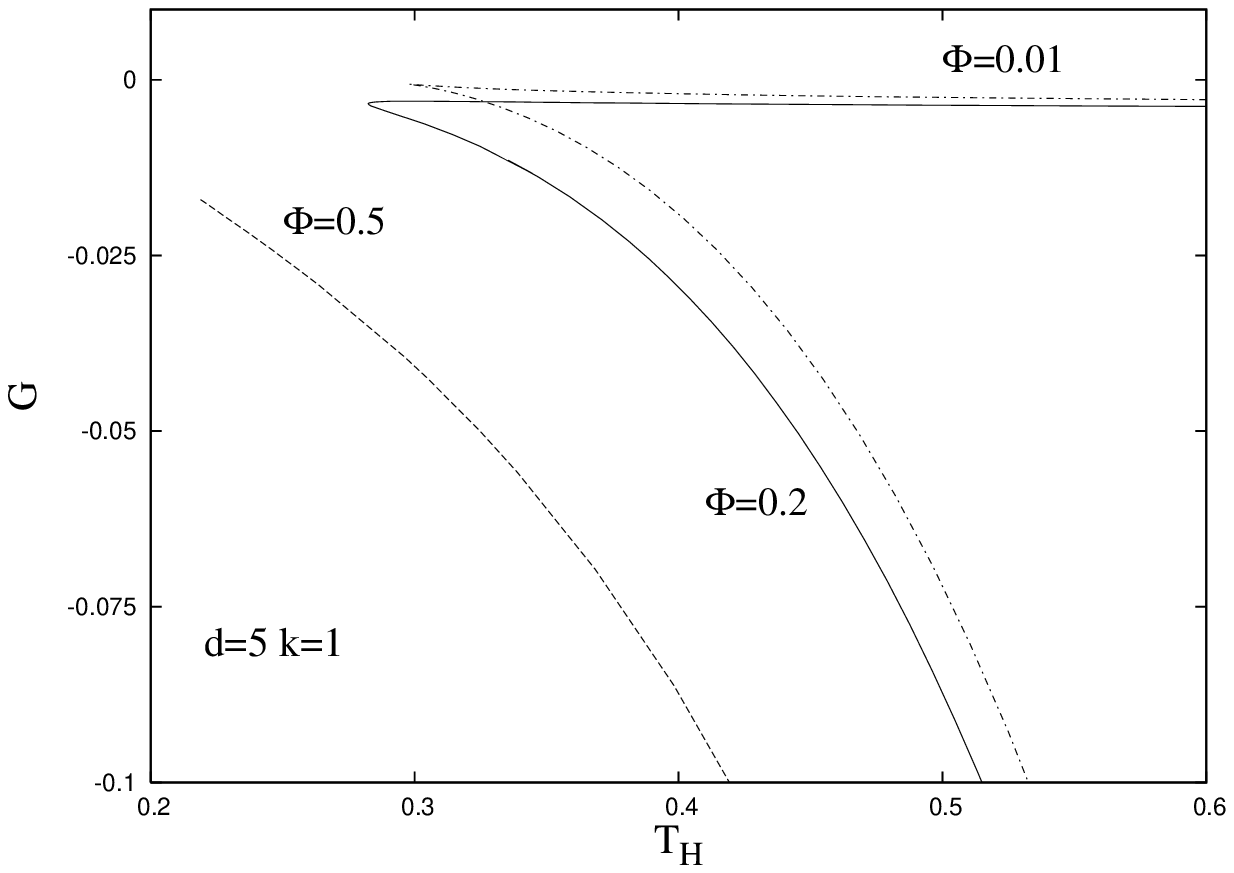}
}}}
\caption{{\small  Left: The inverse temperature $\beta=1/T_H$ ${\it vs.}$ 
the horizon radius $r_h$ for various values of the electric potential $\Phi$.
Right: The Gibbs free energy $G$ ${\it vs.}$ the Hawking temperature $T_H$ 
for charged static black strings in the grand-canonical ensemble.}}
\end{figure}
%
In view of this difficulty one has to resort to an alternative quasi-Euclidean approach.
Using this method we computed in Section $3$ the thermodynamic quantities 
for the electrically charged black string, such as its mass, temperature, entropy, 
charge and electric potential, while in Section $4$ we considered the similar 
thermodynamic quantities for the rotating black strings 
in even dimensions.

Let us consider first the thermodynamics of the charged string. 
As it is well-known, small Schwarzschild-AdS black holes have negative 
specific heat (they are unstable) but large size black holes have
positive specific heat (and they are stable). It was also found 
that Schwarzschild-AdS black holes are subjected to the 
Hawking-Page phase transition \cite{Hawking:1982dh}: at 
low temperatures a thermal gas is a globally stable configuration 
in the AdS background (since in this background, black holes can 
exist only for temperatures above a critical value) but when one 
increases the temperature above the critical value, the thermal 
bath becomes unstable and collapses to form a black hole. 
The results in \cite{Mann:2006yi} indicate that this is the 
case for $k=1$ AdS vacuum black strings as well. The temperature of 
these configurations is bounded from below. At low temperatures
 one has a single bulk solution, corresponding to the thermal 
globally regular solution.

At higher temperatures, above a critical value, 
there exist two bulk solutions that correspond to the so-called small 
(unstable) and large (stable) black string solutions. 
In \cite{Mann:2006yi} it has been suggested that there 
exists a similar Hawking-Page phase transition for the 
neutral black string. The entire unstable branch has 
positive free energy while the free energy of the 
stable branch goes rapidly negative for temperatures 
above a critical value. This critical value corresponds
 then to the temperature at which one observes a phase 
transition between the large black strings and the thermal globally regular background. 

The discussion of the charged case is more involved 
\cite{Chamblin:1999tk}. One can describe the thermodynamics 
using 
two different thermodynamic ensembles. The grand-canonical
 ensemble is defined by coupling the system to energy and charge 
reservoirs at fixed temperature $T_H$ and fixed electric potential 
$\Phi$. As we have seen 
in Section $3$, a computation of the action 
using quasi-Euclidean methods leads to $I=\beta(M-\Phi Q)-S$, such 
that the Gibbs free energy is obtained as 
$G=I/\beta=M-T_HS-\Phi Q.$

One can plot the inverse temperature versus the horizon radius 
as in Figure $6$ (left). Similarly to the RNAdS case one finds that 
there are two types of behaviour, determined by a critical value, 
$\Phi_c$, of the electric potential $\Phi$. For $\Phi<\Phi_c$ one 
finds that the inverse temperature is bounded and goes smoothly 
to zero as $r_h\ra 0$, while for $\Phi\geq\Phi_c$ one finds that 
$\beta$ diverges when the radius of the extremal black string is reached.
A graph of the Gibbs free energy ${\it vs.}$ the Hawking temperature 
is presented in Figure $6$ for various values of the electric potential $\Phi$
(to simplify the picture, we consider only the $d=5,~k=1$ case). 
In the large potential regime one has only one branch of large black 
string solutions and one can see that they are stable. In the low 
potential regime one has two branches of allowed solutions and 
similarly to the RNAdS case the small black strings are unstable, 
while the large black strings are stable and dominate the 
thermodynamics for all temperatures.
%
\begin{figure}[h!]
\parbox{\textwidth}
{\centerline{
\mbox{
\epsfysize=10.0cm
\includegraphics[width=82mm,angle=0,keepaspectratio]{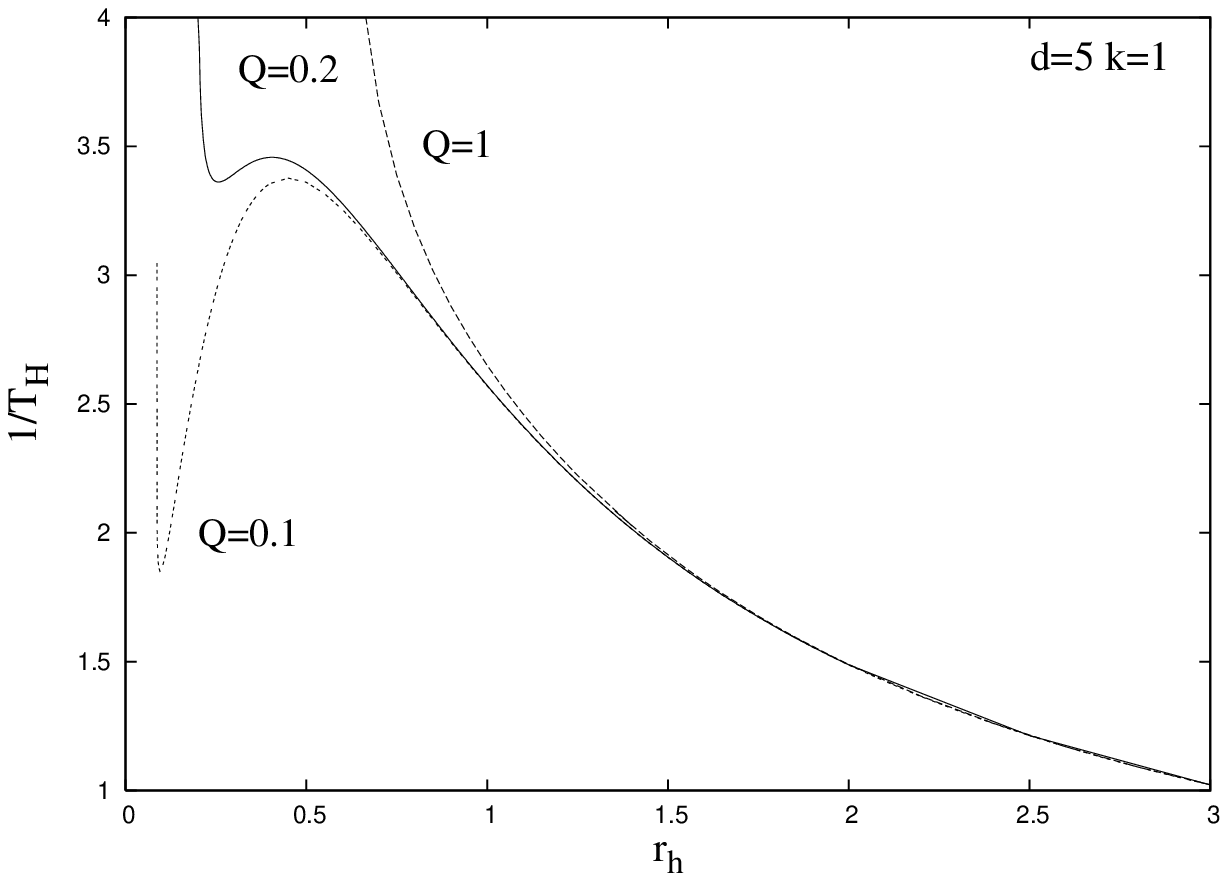}
\includegraphics[width=82mm,angle=0,keepaspectratio]{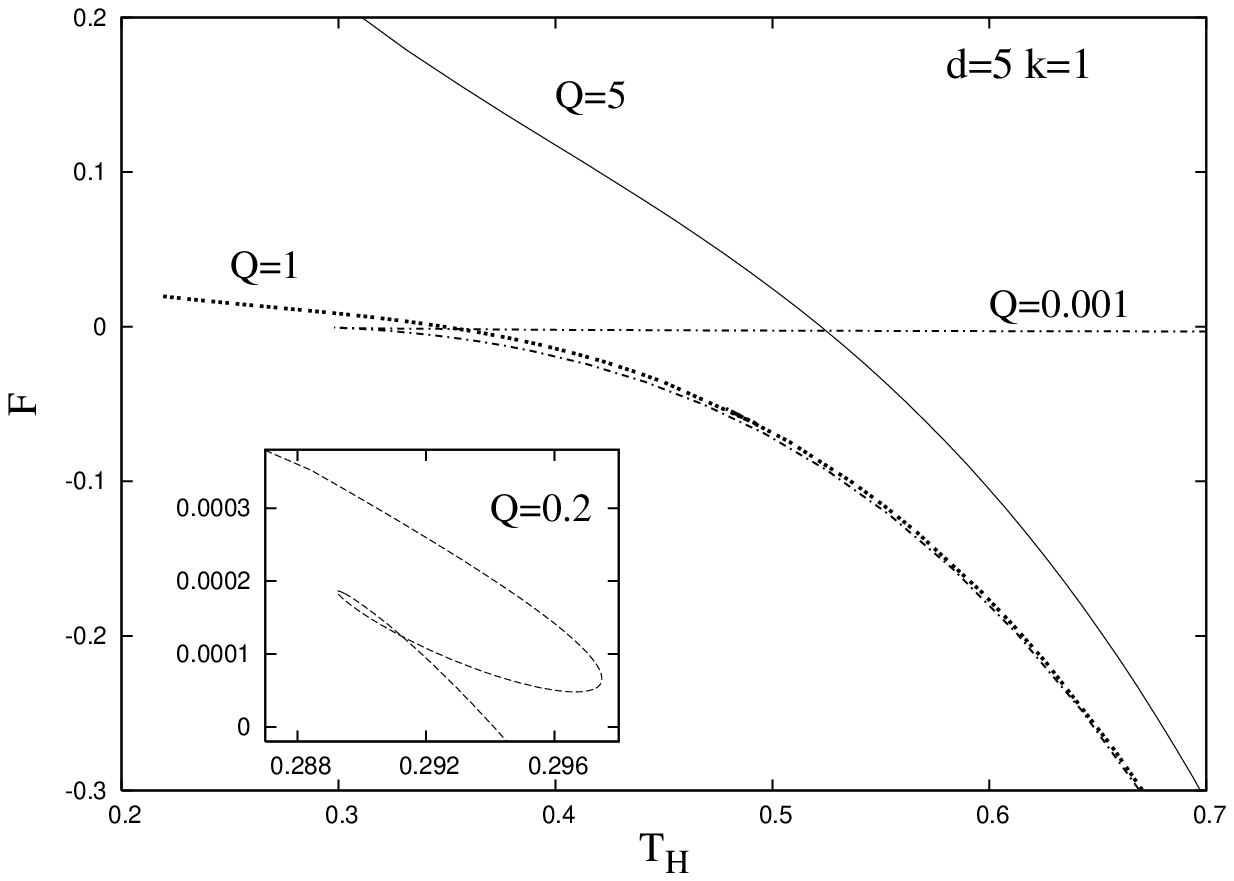}
}}}
\caption{{\small  Left: The inverse temperature $\beta=1/T_H$ ${\it vs.}$ 
the horizon radius $r_h$ for various values of the electric charge $Q$.
Right:  The free energy $F$ ${\it vs.}$ the Hawking temperature $T_H$ 
for charged static black strings in the canonical ensemble.}}
\end{figure}
%
%
If one fixes the value of the charge on the boundary then one
 deals with the canonical thermodynamic 
ensemble. 
The corresponding thermodynamic potential, the 
Helmholtz potential can be obtained from the Gibbs free energy 
by means of a Legendre transformation (see also (\ref{canaction})) 
$F=G+\Phi Q=M-T_HS$.
A plot in Figure $7$ of the inverse temperature $\beta$ $vs.$ 
the black string horizon radius $r_h$ reveals the very interesting 
phase structure found previously in the RNAdS case \cite{Chamblin:1999tk}: 
for small values of the charge, bellow a critical value $Q_{c}$, one finds 
that one can have three branches of black string solutions, while for large charges, 
greater than the critical value, one finds only one branch of large black strings, 
which are thermodynamically stable. 
  
In Figure $7$ we present also a plot of $F$ ${\it vs.}$ temperature for 
several values of $Q$. The corresponding picture for RNAdS black holes was discussed 
in \cite{Chamblin:1999tk}.
The black string picture here resembles again some of 
the features of the RNAdS case. Since there are three black string 
branches for small values of $Q<Q_c$ one can see that the 
free energy presents 
a part resembling the ``swallowtail'' shape of \cite{Chamblin:1999tk} as well.

In the small charge regime, for 
low temperatures, there exists only one branch of black strings 
(``branch $1$''). However, unlike the RNAdS case this branch is unstable. 
Increasing the temperature one notices the apparition of two other 
branches of solutions  (``branch $2$'' and ``branch $3$'') and they 
separate from each other at higher temperatures. 
At some temperature 
the two branches (``branch $1$'' and ``branch $2$'') coalesce and then disappear, 
while ``branch $3$'' carries on till higher temperatures and at high enough 
temperatures the free energy of  ``branch $3$'' goes rapidly negative and, 
therefore, this branch dominates the thermodynamics in this regime. If the 
charge is greater than the critical value, one has only one branch of black 
strings. At high enough temperatures this branch is thermodynamically stable.

The discussion of the thermodynamic properties of the rotating 
black string solutions in even dimensions proceeds on similar lines. 
One has again the choice of two distinct thermodynamic ensembles to 
describe their physics. 
The grand-canonical ensemble corresponds 
in this case 
to fixing the Hawking temperature $T_H$ and the 
angular velocity $\Omega_H$ on the boundary. In this case the 
thermodynamic Gibbs potential is given by $G=M-T_HS-(d-2)\Omega_HJ/2$.
%
\begin{figure}[h!]
\parbox{\textwidth}
{\centerline{
\mbox{
\epsfysize=10.0cm
\includegraphics[width=82mm,angle=0,keepaspectratio]{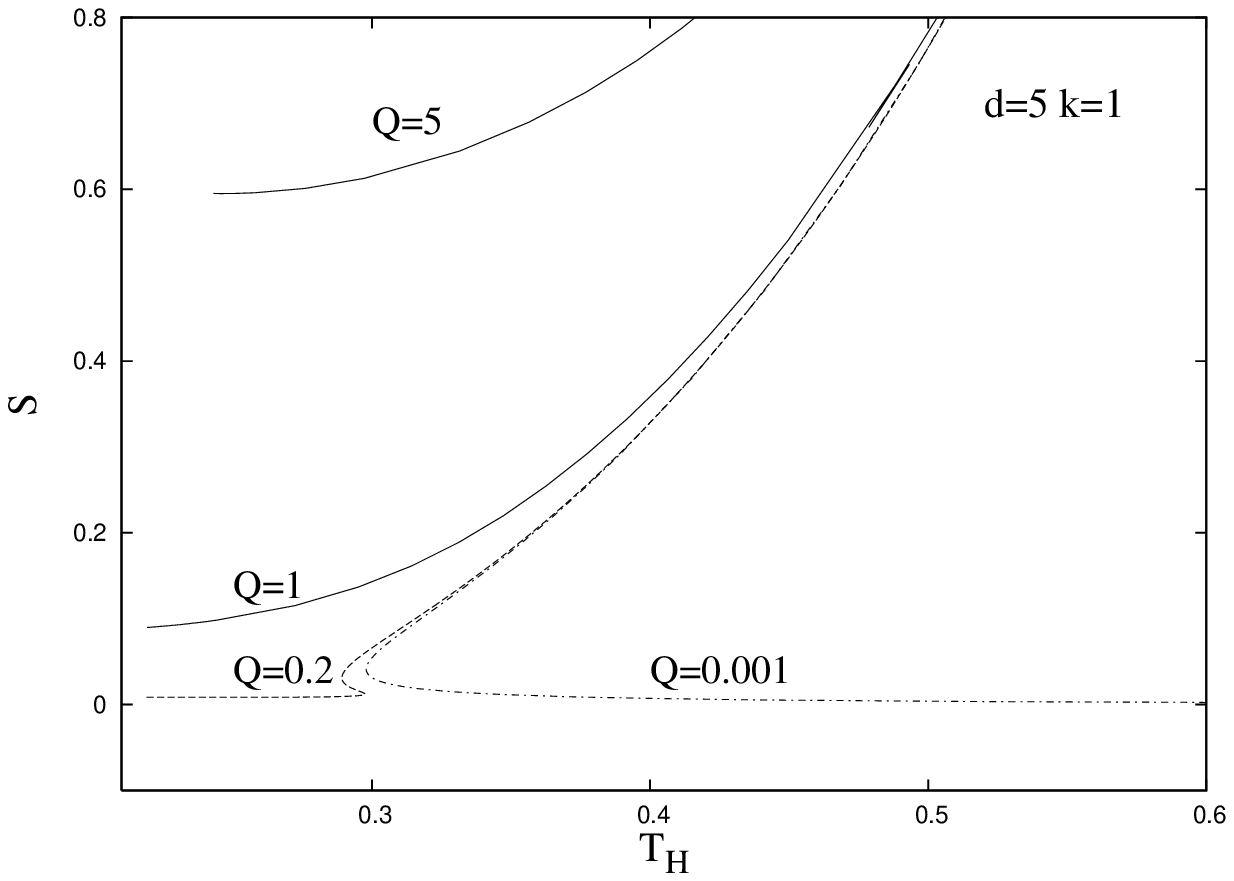}
\includegraphics[width=82mm,angle=0,keepaspectratio]{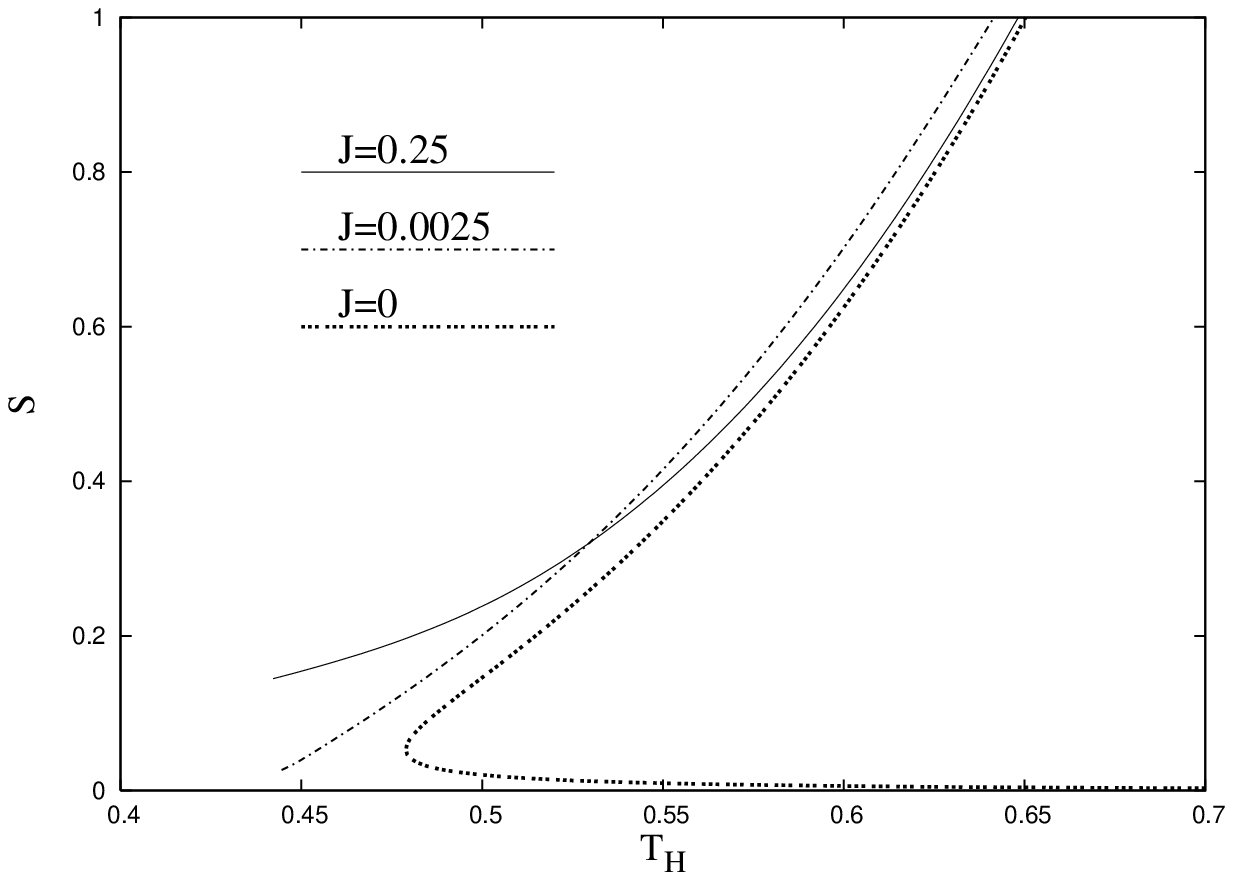}
}}}
\caption{{\small The entropy is plotted \textit{vs.} 
the Hawking temperature for the $k=1$, $d=5$ charged static and the
 $k=1$, $d=6$ vacuum rotating black string solutions.}}
\end{figure}
%
There exists again a critical value of $\Omega_H$: for values 
smaller than this critical value one 
has two branches (one stable and one unstable) of rotating black string solutions, 
while for values 
of $\Omega_H$ greater than the critical value 
one obtains only one branch (unstable) of black string solutions. 
The canonical ensemble corresponds to fixing the Hawking temperature 
$T_H$ and the angular momentum $J$ on the boundary. 
The thermodynamic potential, which in this case it is 
Helmholtz' free energy, can be obtained from the Gibbs 
potential by means of a Legendre transform $F=G+(d-2)\Omega_HJ/2=M-T_HS$. 
In this case our results indicate that the situation familiar from 
the static case remains valid for small values  of $J$.
For larger $J$, one finds however only one branch of solutions, which becomes 
stable at high enough temperatures.

One can also attempt a discussion of the local thermodynamic 
stability of the black string solutions based on the numerical 
results presented in the previous sections, which are valid for 
solutions with Lorentzian signature. Our results indicate that 
the presence of an electric charge or an angular momentum has a 
tendency to stabilize the black strings. Here we shall restrict 
to the case of a canonical ensemble, in which case the length of 
the extra-dimension $L$, the electric charge or the angular momentum 
are kept fixed. The response function whose
 sign determines the thermodynamic stability is the heat capacity
 $C=T\left(\frac{\partial S}{\partial T}\right)_{L,X},$ with $X=Q$ or $J$.
As seen in Figure $8$, the picture familiar from the static
 vacuum case is recovered for small electric charge or a small
 angular momenta. However, our numerical results suggest that 
only one branch of stable solutions exist for large enough 
values of the new global charges.
Along this branch the entropy is increasing with the Hawking 
temperature and, as a result,
the solutions have a positive heat capacity and therefore these 
solutions are locally thermodynamically stable.

\section{Einstein-Maxwell-Liouville black holes in lower dimensions}

Consider a theory in $d$ dimensions described by the 
bosonic Lagrangian:
\beqs
{\cal L}_d&=&\sqrt{|\tilde{g}|}(\tilde{R}-2 \Lambda
-\frac{1}{4} \tilde{F}_{(2)}^2)~,
\label{lagd}
\eeqs
where $\tilde{F}_{(2)}=d\tilde{A}_{(1)}$. 
Assume that the fields are stationary and that the 
system admits two commuting Killing vectors (one of them is 
assumed to be timelike). We will perform a dimensional reduction 
from $d$-dimensions down to $(d-2)$-dimensions along two 
directions $z$ and $t$. Our metric ansatz is:
\beqs
d\tilde{s}_{d}^2&=&e^{-\sqrt{\frac{2(d-3)}{d-2}}\phi}(dz+
{\cal A}_{(1)}^1+{\cal A}_{(0)2}^1dt)^2
\nonumber\\
&&+e^{\sqrt{\frac{2}{(d-2)(d-3)}}\phi}\bigg[-
e^{-\sqrt{\frac{2(d-4)}{d-3}}\phi_1}(dt+
{\cal A}_{(1)}^2)^2+e^{\sqrt{\frac{2}{(d-3)(d-4)}}\phi_1}ds_{d-2}^2\bigg],
\label{KKmetricd}
\eeqs
while the Kaluza-Klein reduction ansatz for the 
electromagnetic potential field is:
\beqs
\tilde{A}_{(1)}=A_{(1)}+A_{(0)2}dt+A_{(0)1}dz.
\eeqs

In these notations, the bosonic Lagrangean in 
$(d-2)$ dimensions, obtained by dimensional reduction 
of the $EM$ theory from $d$-dimensions is given by:
\beqs
{\cal L}_{d-2}&=&\sqrt{|g|}\bigg[R-2\Lambda e^{\sqrt{\frac{2}{(d-2)
(d-3)}}\phi+\sqrt{\frac{2}{(d-3)(d-4)}}\phi_1}-\frac{1}
{2}(\partial\phi)^2-\frac{1}{2}(\partial\phi_1)^2
\nonumber\\
&&-\frac{1}{4} e^{-\sqrt{\frac{2}{(d-2)(d-3)}}\phi-
\sqrt{\frac{2}{(d-3)(d-4)}}\phi_1}(F_{(2)})^2
+\frac{1}{2} e^{\sqrt{\frac{2(d-4)}{(d-3)}}\phi_1-
\sqrt{\frac{2(d-2)}{d-3}}\phi}({\cal F}_{(1)2}^1)^2
\nonumber\\
&&-\frac{1}{4} e^{-\sqrt{\frac{2}{(d-3)(d-4)}}\phi_1
-\sqrt{\frac{2(d-2)}{d-3}}\phi}({\cal F}_{(2)}^1)^2
+\frac{1}{4} e^{-\sqrt{\frac{2(d-3)}{d-4}}\phi_1}
({\cal F}_{(2)}^2)^2\nonumber\\
&&
-\frac{1}{2} e^{\sqrt{\frac{2(d-3)}{(d-2)}}\phi}(F_{(1)1})^2
+\frac{1}{2} e^{-\sqrt{\frac{2}{(d-2)(d-3)}}\phi
+\sqrt{\frac{2(d-4)}{(d-3)}}\phi_1}(F_{(1)2})^2~\bigg].
\eeqs
Notice that the kinetic terms of the field strengths that 
carry an index $2$ (corresponding to the timelike direction 
$t$) have the opposite signs to the usual ones in 
a dimensional reduction along a spatial direction. 
In general, the field strengths that appear in the 
dimensionally reduced Lagrangian have Chern-Simons type 
modifications and can be expressed in the following form:
\beqs
{\cal F}_{(1)2}^1&=&d{\cal A}_{(0)2}^1\equiv d\chi,~~~~
{\cal F}_{(2)}^1={\cal A}_{(1)}^1-d\chi\wedge{\cal A}_{(1)}^2,~~~~
{\cal F}_{(2)}^2=d{\cal A}_{(1)}^2,~~~~
F_{(1)1}=dA_{(0)1}\nonumber\\
F_{(1)2}&=&dA_{(0)2}-\chi dA_{(0)1},~~~~
F_{(2)}=dA_{(1)}-dA_{(0)1}\wedge({\cal A}_{(1)}^1
-\chi {\cal A}_{(1)}^2)-dA_{(0)2}\wedge {\cal A}_{(1)}^2~.{~~~~}
\eeqs
We shall prove now that the above Lagrangian has a global symmetry group $SL(2,R)/O(1,1)$.

First we perform the following field redefinitions:
\beqs
{\cal A}_{(1)}^1&=&\hat{{\cal A}}_{(1)}^1+\chi \hat{{\cal A}}_{(1)}^2,~~~
{\cal A}_{(1)}^2=\hat{{\cal A}}_{(1)}^2,~~~
A_{(1)}=\hat{A}_{(1)}+A_{(0)1}\wedge \hat{{\cal A}}_{(1)}^1+A_{(0)2}
\wedge \hat{{\cal A}}_{(1)}^2
\eeqs
and in terms of the hatted potentials we obtain:
\beqs
{\cal F}_{(2)}^1&=&d\hat{{\cal A}}_{(1)}^1+\chi d\hat{{\cal A}}_{(1)}^2,~~~~~~~
{\cal F}_{(2)}^2=d\hat{{\cal A}}_{(1)}^2,
\eeqs 
respectively
\beqs
F_{(2)}&=&d\hat{A}_{(1)}+A_{(0)1}\wedge d\hat{{\cal A}}_{(1)}^1
+A_{(0)2}\wedge d\hat{{\cal A}}_{(1)}^2 \nonumber\\
&=&d\hat{A}_{(1)}+\left(\begin{array}{cc}A_{(0)2}&A_{(0)1}
\end{array}\right)\left(\begin{array}{cc}0&1\\1&0\end{array}
\right)\left(\begin{array}{c}\hat{{\cal F}}_{(2)}^1\\\hat{{\cal F}}_{(2)}^2
\end{array}\right),
\eeqs
where $\hat{{\cal F}}_{(2)}^i=d\hat{{\cal A}}_{(1)}^i$, with $i=1, 2$. We make a further rotation of the scalar fields:
\beqs
\left(\begin{array}{c}\hat{\phi}\\\hat{\phi_1}\end{array}\right)&=&
\left(\begin{array}{cc}\sqrt{\frac{d-4}{2(d-3)}}&\sqrt{\frac{d-2}{2(d-3)}}
\\\sqrt{\frac{d-2}{2(d-3)}}&-\sqrt{\frac{d-4}{2(d-3)}}\end{array}\right)
\left(\begin{array}{c}\phi \\\phi_1\end{array}\right),
\label{rotsc}
\eeqs
and define the matrix ${\cal M}$ by:
\beqs {\cal M}&=&\left(\begin{array}{cc}e^{-\hat{\phi}_1}&\chi 
e^{-\hat{\phi}_1}\\\chi e^{-\hat{\phi}_1}&-e^{\hat{\phi}_1}
+\chi^2e^{-\hat{\phi}_1}\end{array}\right).\label{M}\eeqs
Note that $\det{\cal M}=-1$ hence ${\cal M}$ is not an $SL(2,R)$ matrix. The $(d-2)$-dimensional Lagrangean can then be written in the following compact form:
\beqs
{\cal L}_{d-2}&=&\sqrt{|g|}(R-2 \Lambda e^{\frac{2\hat{\phi}}
{\sqrt{(d-2)(d-4)}}}-\frac{1}{2} (\partial\hat{\phi})^2
+\frac{1}{4} \tr[\partial{\cal
M}^{-1}\partial{\cal M}]-\frac{1}{4} e^{-\frac{2\hat{\phi}}
{\sqrt{(d-2)(d-4)}}}(F_{(2)})^2\nonumber\\
&&-\frac{1}{4} e^{-\sqrt{\frac{d-2}{d-4}}\hat{\phi}} 
\hat{{\cal F}}_{(2)}^T{\cal M}\hat{{\cal F}}_{(2)}-
\frac{1}{2} e^{\sqrt{\frac{d-4}{d-2}}\hat{\phi}}F_{(1)}^T({\cal M}^{-1})F_{(1)})~,
\eeqs
where we have defined the column vectors:
\beqs F_{(1)}=\left(\begin{array}{c}dA_{(0)1}
\nonumber\\ 
dA_{(0)2}\end{array}\right),~~~~~~~~
\hat{{\cal F}}_{(2)}=\left(\begin{array}{c}\hat{{\cal F} }_{(2)}^1
\\\hat{{\cal F}}_{(2)}^2\end{array}\right).\eeqs

This Lagrangian is manifestly invariant under $SL(2,R)$ 
transformations if we consider the following transformation laws for the potentials:
\beqs
g_{\mu\nu}&\ra& g_{\mu\nu},~~~~~~~ \Lambda\ra \Lambda,
~~~~~~~{\cal M}\rightarrow ~\Omega^T{\cal M}\Omega,
~~~~~~~\hat{\phi}\rightarrow \hat{\phi},
~~~~~~~\hat{A}_{(1)}\rightarrow \hat{A}_{(1)}
\nonumber\\
\left(\begin{array}{c}\hat{{\cal A}}_{(1)}^1
\nonumber\\\hat{{\cal A}}_{(1)}^2\end{array}
\right)&\rightarrow& \Omega^{-1}
\left(\begin{array}{c}\hat{{\cal A}}_{(1)}^1
\nonumber\\\hat{{\cal A}}_{(1)}^2\end{array}\right),~~~~~
\left(\begin{array}{c}A_{(0)1}\nonumber\\
A_{(0)2}\end{array}\right)\rightarrow 
\Omega^T\left(\begin{array}{c}A_{(0)1}\nonumber\\ A_{(0)2}
\end{array}\right),
\label{sltransform}
\eeqs
where $\Omega\in SL(2,R)$.

The ansatz (\ref{lagd})-(\ref{KKmetricd}) generalizes the transformation considered 
previously in \cite{Mann:2006yi} by the inclusion of 
an electromagnetic field in the initial Lagrangian in 
$d$ dimensions and, also, by considering a general 
stationary metric ansatz in $(d-1)$ dimensions
(see also \cite{Charmousis:2006fx}
for other work in this direction). 
Therefore, it is appropriate when treating the charged black string solutions as well 
as the neutral rotating solutions.

\subsection{The electrically charged solutions}
Let us apply now this technique to the new solutions 
presented in this paper. Consider first the case of 
the electrically charged black strings described in 
Section $3$. Starting with the $d$ dimensional metric 
(\ref{metric}) and electromagnetic field (\ref{Vr}) 
and performing a double dimensional reduction along 
the coordinates $z$ and $t$ we obtain in $(d-2)$ dimensions the following fields:
\beqs
ds_{d-2}^2&=&\left(a b\right)^{\frac{1}{d-4}}
\left(\frac{dr^2}{f}+r^2d\Sigma^2_{k,d-3}\right),
~~~~~~~e^{-\sqrt{\frac{2(d-3)}{d-2}}\phi}=a,~~~~~~~
e^{-\sqrt{\frac{2(d-4)}{d-3}}\phi_1}=ba^{\frac{1}{d-3}},\nonumber\\
{\cal A}_{(1)}^1&=&0,~~~{\cal A}_{(1)2}^1\equiv\chi=0,
~~~{\cal A}_{(1)}^2=0,~~~A_{(1)}=0,~~~A_{(0)2}=2V(r),~~~A_{(0)1}=0.
\eeqs
Let us parameterize the matrix $\Omega$ in the form:
\beqs \Omega=\left(\begin{array}{cc} \alpha & \beta 
\\  \gamma & \delta \\\end{array}\right)
, \qquad \alpha\delta-\beta\gamma=1.
\eeqs
Rotating the scalar fields in $(d-2)$ dimensions and 
acting with the $SL(2,R)$ transformation $\Omega$ on the
 corresponding matrix ${\cal M}$, the final scalar fields read:
\beqs
e^{\phi}&=&(\alpha^2 a-\gamma^2 b)^{-\sqrt{\frac{d-2}{2(d-3)}}},~~~~~~~
e^{\phi_1}=(a b)^{-\sqrt{\frac{d-3}{2(d-4)}}}(\alpha^2 a-
\gamma^2 b)^{-\sqrt{\frac{d-4}{2(d-3)}}},
\eeqs
while the final electromagnetic potentials are given by:
\beqs
A_{(0)1}=2\gamma V(r),~~~~~~~A_{(0)2}=2\delta V(r).
\eeqs
Gathering all these results, we obtain in $(d-1)$ dimensions 
the following fields:
\beqs
 ds_{d-1}^2&=&-ab(\alpha^2 a-\gamma^2 b)^{-\frac{d-4}{d-3}}dt^2+
(\alpha^2 a-\gamma^2 b)^{\frac{1}{d-3}}\frac{dr^2}{f}+r^2
(\alpha^2 a-\gamma^2 b)^{\frac{1}{d-3}}d\Sigma^2_{k,d-3},
\\
 e^{\phi}&=&(\alpha^2 a-\gamma^2 b)^{-\sqrt{\frac{d-2}{2(d-3)}}},~~~~~
{\cal A}_{(1)}=\frac{\alpha\beta a-\gamma\delta b}{\alpha^2 
a-\gamma^2 b}dt,~~~~~A_{(0)1}=2\gamma V(r),~~~~~A_{(0)2}=2\delta V(r).
\nonumber
\label{finel}
\eeqs
For $F_{(1)1}=dA_{(0)1}$ and $F_{(2)}=dA_{(0)2}\wedge dt$ this 
gives us a solution of the equations of motion derived from the 
following Lagrangian:
\beqs
{\cal L}_{d-1}&=&\sqrt{|g|}(R-2 \Lambda e^{\sqrt{\frac{2}
{(d-2)(d-3)}}\phi}-\frac{1}{2}e(\partial\phi)^2-
\frac{1}{2} e^{\sqrt{\frac{2(d-3)}{(d-2)}}\phi}
(F_{(1)1})^2 \nonumber\\
&&-\frac{1}{4} e^{-\sqrt{\frac{2}{(d-2)(d-3)}}\phi}
(F_{(2)})^2-\frac{1}{4} e^{-\sqrt{\frac{2(d-2)}{d-3}}
\phi}({\cal F}_{(2)})^2)~,
\label{Ld-1el}
\eeqs
which corresponds to an Einstein-Dilaton theory with 
a Liouville potential for the dilaton and coupled with 
two electromagnetic fields and one scalar field. 

\subsection{The rotating solutions}
Take now the case of the rotating neutral black string 
solutions in even $d$ dimensions. One can set directly 
$A_{(1)}=A_{(0)1}=A_{(0)2}=0$ in the initial ansatz for 
the $d$ dimensional fields and by comparing the metric 
(\ref{metric-rot}) with the Kaluza-Klein ansatz (\ref{KKmetricd}) 
one reads straightforwardly the following fields in $(d-2)$ dimensions:
\beqs
ds_{d-2}^2&=&\left(a(b-r^2w^2)\right)^{-\frac{1}{2}
\sqrt{\frac{d-2}{d-4}}}\bigg[\frac{dr^2}{f}+
g\sum_{i=1}^{N-1}\left(\prod_{j=0}^{i-1} \cos^2\theta_j \right) 
d\theta_i^2+g\sum_{k=1}^N \rho_kd\vphi_k^2\cr\nonumber\\
&& 
  +\frac{r^4(1-w^2)+r^2(b+gw^2)-bg}{b-r^2w^2}
\left(\sum_{k=1}^N\rho_kd\vphi_k\right)^2\bigg],\\
 e^{-\sqrt{\frac{2(d-3)}{d-2}}\phi}&=&a,~~
e^{-\sqrt{\frac{2(d-4)}{d-3}}\phi_1}=
(b-r^2w^2)a^{\frac{1}{d-3}},~~{\cal A}_{(1)}^2
=\frac{r^2\sum_{k=1}^N\rho_kd\vphi_k}{b-r^2w^2},~~{\cal A}_{(1)}^1=0,
~~{\cal A}_{(0)2}^1=0,\nonumber
   \eeqs
where we denoted:
\beqs
\rho_k&=&\left( \prod_{l=0}^{k-1} \cos^2 
\theta_l\right) \sin^2\theta_k.
\eeqs
Rotating the scalar fields in $(d-2)$ dimensions 
and acting with the $SL(2,R)$ transformation $\Omega$ 
on the corresponding matrix ${\cal M}$, the final scalar fields read:
\beqs
e^{\phi}&=&\left(\alpha^2 a-\gamma^2 (b-r^2w^2)
\right)^{-\sqrt{\frac{d-2}{2(d-3)}}},~  
e^{\phi_1}=\big[a (b-r^2w^2)\big]^{-
\sqrt{\frac{d-3}{2(d-4)}}}\left(\alpha^2 a-\gamma^2 (b-r^2w^2)
\right)^{-\sqrt{\frac{d-4}{2(d-3)}}}\nonumber,
\eeqs
while the final electromagnetic potentials are given by:
\beqs
{\cal A}_{(1)}^1=(\alpha\chi-\beta) \frac{r^2
\sum_{k=1}^N\rho_kd\vphi_k}{b-r^2w^2},~~~
{\cal A}_{(1)}^2=\alpha \frac{r^2\sum_{k=1}^N\rho_kd\vphi_k}{b-r^2w^2},
\eeqs
where
\beqs
\chi\equiv {\cal A}_{(0)2}^1=\frac{\alpha\beta a-
\gamma\delta (b-r^2w^2)}{\alpha^2 a-\gamma^2 (b-r^2w^2)}.
\eeqs
Gathering all these results, we obtain in $(d-1)$ 
dimensions the following fields:
\begin{eqnarray}
ds_{d-1}^2=-a(b-r^2w^2)\big[\alpha^2 a-\gamma^2 
(b-r^2w^2)\big]^{-\frac{d-4}{d-3}}(dt+{\cal A}_{(1)}^2)^2
+\big[\alpha^2 a-\gamma^2 (b-r^2w^2)\big]^{\frac{1}{d-3}}ds_{d-2}^2,~~~~~~{~~~}
\\
e^{\phi}=\left(\alpha^2 a-\gamma^2 (b-r^2w^2)
\right)^{-\sqrt{\frac{d-2}{2(d-3)}}},~
{\cal A}_{(1)}=\frac{\alpha\beta a-\gamma\delta 
(b-r^2w^2)}{\alpha^2 a-\gamma^2 (b-r^2w^2)}dt+
(\alpha\chi-\beta) \frac{r^2\sum_{k=1}^N\rho_kd\vphi_k}{b-r^2w^2}.~~{~~~}
\nonumber
\label{final}
\end{eqnarray}
This gives us a solution of the equations of motion 
derived from the following Lagrangian:
\beqs
{\cal L}_{d-1}&=&\sqrt{|g|}(R-2 \Lambda e^{\sqrt{\frac{2}
{(d-2)(d-3)}}\phi}-\frac{1}{2} (\partial\phi)^2-
\frac{1}{4} e^{-\sqrt{\frac{2(d-2)}{d-3}}\phi}({\cal F}_{(2)})^2)~,
\label{Ld-1}
\eeqs
which corresponds to an EM-Dilaton 
theory with a Liouville potential for the dilaton.

As a consistency check of our final solutions, 
one can see that if one takes $\Omega=I_2$ then 
one obtains the initial charged black string solution 
(\ref{metric}), respectively the neutral rotating
 black string solution (\ref{metric-rot}) after 
oxidizing back to $d$ dimensions. 
Also, if $\alpha=\delta=\cosh p$ and $\beta=\gamma=\sinh p$ 
the effect of the $SL(2,R)$ transformation is equivalent 
to a boost of the initial black string solution in the $z$ direction.

Let us notice that the final solution (\ref{final}) 
corresponds to a rotating charged black hole solution 
of the Einstein-Maxwell-Dilaton theory with a Liouville 
potential for the dilaton, generalizing the static charged 
black hole solution derived previously in \cite{Mann:2006yi}, 
which is recovered from the above by setting $w=0$. For a 
generic Kaluza-Klein dimensional reduction, 
if the isometry generated by the Killing vector 
$ \partial/\partial z$ has fixed points then the 
dilaton $\phi$ will diverge and the $(d-1)$-dimensional metric 
will be singular at those points. However, this is not the case 
for our initial black string solutions and therefore the 
$(d-1)$-dimensional fields are non-singular in the near-horizon limit $r\ra r_h$. 
Indeed, 
in the near horizon limit $b(r)\ra 0$ while $a(r)\ra a_0$ 
and the $(d-1)$-dimensional fields are non-singular. As with 
the black hole solution discussed in \cite{Mann:2006yi}, the 
situation changes when we look in the asymptotic region. 
Recall from (\ref{KKmetricd}) that 
\begin{eqnarray}
g_{zz}\equiv e^{-\sqrt{\frac{2(d-3)}{d-2)}}\phi}=\alpha^2 a-\gamma^2 (b-r^2w^2)
\end{eqnarray}
gives the radius squared of the $z$-direction in $d$-dimensions and that it 
diverges in the large $r$ limit. 
Then for generic values of the parameters in 
$\Omega$ we find that $g_{zz}\sim r^2$ and the dilaton
 field in $(d-1)$-dimensions will diverge in the asymptotic region. 
Physically, this means that the spacetime decompactifies at infinity; 
the higher-dimensional theory should be used when describing 
such black holes in these regions. On the other hand one can 
choose the parameters in $\Omega$ such that $\alpha=\gamma$. 
In this case the  asymptotic behaviour of 
the $(d-1)$-dimensional fields is quite different as 
$g_{zz}\sim \frac{c_z-c_t}{r^{d-3}}$ and the radius of 
the $z$-direction collapses to zero asymptotically. 
However the dilaton field still diverges at infinity.
Also, one can prove that the relevant properties of these solutions
can be derived from the corresponding $d-$dimensional seed configurations.

\subsection{$T$-dual solutions in $10$ dimensions}
Finally, let us make a few remarks about the embedding of 
our solutions into string/M-theory. 
Our neutral, rotating black string solutions were being 
considered in even dimensions only and, therefore, 
since they admit a negative cosmological constant,
 one cannot give them a natural interpretation as solutions 
of string/M-theory. However, the charged $5$-dimensional 
black string solution can be given such an interpretation. 
To see this, consider the oxidation back to $d=5$ dimensions of 
the solution given in (\ref{finel}):\footnote{We normalized the 
electromagnetic field such that the action in $5$ dimensions 
has again the form (\ref{action}).}
\beqs
ds_5^2&=&-\frac{ab}{\alpha^2 a-\gamma^2 b}dt^2+(\alpha^2 a-
\gamma^2 b)(dz+\frac{\alpha\beta a-\gamma\delta b}{\alpha^2 a-
\gamma^2 b}dt)^2+\frac{dr^2}{f}+r^2d\Omega^2,\\
A_{(1)}&=&\delta V(r) dt+\gamma V(r) dz.
\label{5del}
\eeqs 
Notice that this solution can be obtained from that of 
the initial charged black string by making the following 
coordinate transformation:
\beqs
\left(\begin{array}{c}z\nonumber\\
t\end{array}\right)\rightarrow \Omega\left(\begin{array}{c}z
\nonumber\\ 
t\end{array}\right).
\eeqs
Now, even if the two metrics can be locally isometric 
by means of the above transformation, their global 
structure can be completely different since the coordonate 
transformation considered above mixes the time coordinate $t$ 
with a periodic spacelike coordinate $z$. This is also apparent 
from the above discussion in which we have seen that for special 
values of the parameters in $\Omega$ the asymptotic radius of the 
$z$-direction differs considerably: in some cases it can grow like $r^2$,
 while in others it can collapse to zero at large distances.

In \cite{Chamblin:1999tk} it was shown how to obtain the $5$-dimensional 
EM theory with a negative cosmological constant from a 
Kaluza-Klein reduction of Type IIB supergravity on a $5$-sphere $S^5$. 
The reduction ansatz is given by:
\beqs
ds^2&=&ds_5^2+\ell^2\sum_{i=1}^3\bigg[d\mu_i^2
+\mu_i^2\left(d\varphi_i+\frac{2}{\sqrt{3}\ell}A_{(1)}\right)^2\bigg],
\eeqs
where $ds_5^2$ is the $5$ dimensional metric, $\mu_i$ are 
direction cosines on the $S^5$ (with $\sum_{i=1}^3\mu_i^2=1$) 
and $\varphi_i$ are rotation angles on $S^5$. 
The ansatz for the Ramond-Ramond $5$-form was also 
presented in \cite{Chamblin:1999tk}. 
One can see that a non-vanishing electromagnetic 
field $A_{(1)}$ in $5$ dimensions corresponds to 
a rotation of the $S^5$ by equal amounts in each of its 
three independent rotation planes. Notice now that a 
point at fixed $\mu_i$ and $\varphi_i$ on the $S^5$ moves 
on an orbit of $\xi=\partial/\partial t$. 
For the electric black string solution discussed 
in Section $3$ the norm of this Killing 
vector is:\footnote{We use a gauge in which the electromagnetic 
potential is regular on the horizon.} 
\beqs
\xi^{\mu}\xi_{\mu}=-b(r)+\frac{4}{3}V(r)^2.
\eeqs
As explained in \cite{Hawking:1999dp}, 
the existence of regions where the norm of 
this vector becomes positive would signal 
potential instabilities of this metric since 
in those regions the internal sphere $S^5$ would 
rotate at speeds higher than the velocity of light. 
Similarly to the RNAdS case discussed in 
\cite{Hawking:1999dp}, we find that the norm of this 
Killing vector is always negative and, therefore, 
the internal sphere $S^5$ never rotates at speeds faster than light.

It is also instructive to note that since 
the black string horizon is translationally 
invariant along the $z$ direction, 
one can consider the effect of a $T$-duality 
along this spacelike direction. 
We shall focus here on the specific $T$-duality 
that interchanges the Type IIA with the Type IIB
supergravity solutions. The action of $T$-duality on the supergravity fields 
is easy to write down in
the $NS-NS$ sector. However in the $RR$ sector 
the situation is more
complicated as the $RR$ fields transform in the 
spinorial representation of the $T$-duality group \cite{Hassan:1999bv,Hassan:1999mm}. 
For a single $T$-duality transformation, say along 
the spacelike coordinate $z$, in the $NS-NS$ sector we 
have the following transformation laws \cite{buscher,Bergshoeff:1995as}:
\beqs 
\tilde{G}_{zz}&=&\frac{1}{G_{zz}},~~~~
\tilde{G}_{zi}=\frac{B_{zi}}{G_{zz}},~~~~
\tilde{G}_{ij}=G_{ij}-\frac{G_{zi}G_{zj}-B_{zi}B_{zj}}{G_{zz}},
\nonumber\\
\tilde{B}_{zi}&=&\frac{G_{zi}}{G_{zz}},~~~~
\tilde{B}_{ij}=B_{ij}-\frac{G_{zi}B_{jz}-G_{zj}B_{iz}}{G_{zz}},~~~~
e^{2\tilde{\phi}}=\frac{e^{2\phi}}{G_{zz}},\label{buscher}
\eeqs
while in the $RR$ sector we have the following transformations laws of the
field strengths $F$ \cite{Hassan:1999bv}:
\beqs 
\tilde{F}^{(n)}_{i_1\cdots i_{n-1}z}&=&F^{(n-1)}_{i_1\cdots
i_{n-1}}-(n-1)G_{zz}^{-1}F^{(n-1)}_{[i_1\cdots
i_{n-2}z}G_{i_n]z},\nonumber\\
\tilde{F}^{(n)}_{i_1\cdots
i_n}&=&F^{(n+1)}_{i_1\cdots i_nz}+nF^{(n-1)}_{[i_1\cdots
i_{n-1}}B_{i_n]z}+n(n-1)G_{zz}^{-1}F^{(n-1)}_{[i_1\cdots
i_{n-2}z}B_{i_{n-1}z}G_{i_n]z}~,
\label{10dRR}
\eeqs
where the indices $i$, $j$ run over all the coordinates except $z$, the
anti-symmetrization does not involve the index $z$ and the upper index for the
$RR$ field strengths $F^{(n)}$ indicates their rank, with $n=0, 2, 4, 6, 8,
10$ for Type IIB theory, while $n=1, 3, 5, 7, 9$ for Type IIA theory.
However, we notice that these field strengths are not independent and they are
dual to each other. In consequence the effective action of the respective
Type II theory contains only field strengths with $n\leq 4$.

For simplicity, we consider the $10$ dimensional oxidation of the 
`boosted' neutral black string solution (\ref{5del}). 
The corresponding ten-dimensional solution of the 
Type IIB supergravity theory is given by:
\beqs 
ds_{10}^2&=&-\frac{ab}{\alpha^2 a-\gamma^2 b}dt^2
+(\alpha^2 a-\gamma^2 b)(dz+\frac{\alpha\beta a
-\gamma\delta b}{\alpha^2 a-\gamma^2 b}dt)^2+\frac{dr^2}{f}+r^2d\Omega^2
+l^2d\Omega_5^2,\nonumber
\eeqs
where $d\Omega_5^2$ is the metric element of the unit $5$-sphere. This metric is a
solution of the Type IIB supergravity equations of motion with the self-dual
$5$-from given by \cite{Chamblin:1999tk}:
\beqs 
F^{(5)}&=&-\frac{4}{\ell}\epsilon_{(5)}-\frac{4}{\ell}\star\epsilon_{(5)}.
\eeqs
On components one has:
\beqs 
F_{\mu_1\mu_2\mu_3\mu_4\mu_5}&=&
\frac{4}{\ell}\epsilon_{\mu_1\mu_2\mu_3\mu_4\mu_5},~~~~
F_{m_1m_2m_3m_4m_5}= -\frac{4}{\ell}\epsilon_{m_1m_2m_3m_4m_5}.
\eeqs
Here the indices $\mu$ are indices on the $AdS_5$ sector, while the $m$'s
correspond to the indices of the coordinates on the $5$-sphere. Our definition
of the Hodge operator $\star$ is such that 
$\star^2F^{(n)}=(-1)^{n+1}F^{(n)}$. 
Using (\ref{buscher}) and (\ref{10dRR}) one finds:
\beqs
ds_{10}^2&=&-\frac{ab+\left(\alpha\beta a-\gamma\delta b\right)^2}
{\alpha^2 a-\gamma^2 b}dt^2+\frac{dz^2}{\alpha^2 a-\gamma^2 b}
+\frac{dr^2}{f}+r^2d\Omega^2+l^2d\Omega_5^2,\nonumber\\
B_{zt}&=&\frac{\alpha\beta a-\gamma\delta b}{\alpha^2 a-\gamma^2 b},
~~~~e^{-2\phi}=\alpha^2 a-\gamma^2 b,
\\
\nonumber
F^{(4)}_{\mu_1\cdots\mu_4}&=&\frac{4 }{\ell}\sqrt{\alpha^2 a-\gamma^2 b}
~\epsilon_{\mu_1\cdots\mu_4},~~~~
F^{(6)}_{i_1\cdots i_6}=-\frac{4}{\ell}\sqrt{\alpha^2 a-\gamma^2 b}
~\epsilon_{i_1\cdots\i_6}
\eeqs
where in the above formulae 
$\mu\neq z$ and $i\neq t, r, \theta, \varphi$,
 while $\epsilon$ denotes the volume element of the corresponding subspace. 
It is an easy matter to check that $\star F^{(4)}=F^{(6)}$ 
while $\star F^{(6)}=-F^{(4)}$ as expected.

Similarly to the case of the $T$-dual geometry of a toroidal black 
hole described in \cite{Rinaldi:2002tc} one 
finds that the horizon location of the $T$-dual 
geometry is the same, given by the root $r=r_h$ of the equation $f(r)=0$. 
The asymptotic limit is non-trivial:
\beqs
ds_{10}^2&\sim&\frac{r^2}{\ell^2}\big[-dt^2
+\ell^2d\Omega_2^2\big]+\frac{\ell^2}{r^2}\big[dr^2+dz^2]+\ell^2d\Omega_5^2,
\eeqs
for generic values of the parameters describing 
the $SL(2,R)$ transformation $\Omega$. 
Apart from the metric element of the $5$-sphere $S^5$ one notices 
that the topology at large $r$ is a warped product of a $3$-dimensional 
Einstein universe with a $2$-dimensional hyperbolic space 
(compactified along the $z$ direction). 
However, for the special values of 
the parameters in $\Omega$, such that 
$\alpha=\gamma$, one finds that asymptotically 
the radius of the $z$ direction grows like $r^2$ and 
therefore the topology of the $5$-dimensional 
part of the metric is very similar to the topology 
of the initial black string solution.

\section{Summary and Discussion}
In this paper we have presented arguments 
for the existence of static electrically charged as 
well as neutral rotating black string-type solutions of 
Einstein gravity with negative cosmological constant.
These are the AdS natural counterparts of the 
$\Lambda=0$ uniform black string solutions.
The first configuration, discussed in Section $3$, 
generalized the neutral black string configurations of 
Ref. \cite{Copsey:2006br,Mann:2006yi} by coupling them 
to an electromagnetic field.  In Section $4$ we described a 
rotating generalization of the black strings in even dimensions only, 
with equal angular momenta. 
Different from the neutral case, no vortex-type solution
was found in these cases.
A  discussion of the thermodynamical 
features of the black string solutions
was presented in Section $5$.
It is interesting to note that in all considered cases (inclusing the
static neutral solutions in \cite{ Mann:2006yi}), the
thermodynamics of the black strings appears to follow the pattern of the 
corresponding black hole solutions with spherical horizons in AdS backgrounds.

On general grounds, one expects the black objects 
with a boundary topology  $R\times S^{d-3}\times S^1$ 
to present a richer phase structure
than in the usual $R\times S^{d-3} $ case (with $R$ corresponding to the time direction), 
as the
radius of the circle introduces another macroscopic scale in the 
theory\footnote{ Charged bubble solutions can be constructed by taking  
  the analytic continuation $z \to i u $, $t \to i\xi$ 
in the general line element (\ref{metric})  
$ds^2=-\bar a(r)du^2+\bar b(r)d\xi^2+ \frac{dr^2}{\bar f(r)}
+r^2d\Sigma^2_{k,d-3},$
(where $\xi$ has a periodicity $\beta$ 
fixed by the absence of conical 
singularities in the $(r,\xi)$ sector of the metric), while $A=\bar V(r)du$.
The field equations can be solved by using the methods in Section 3,
with a very similar set of boundary conditions. 
As emphasized already, one cannot use the numerical solutions in Section 4 to
derive the features of the charged bubble solutions. This is clearly seen
by considering the $k=0$ solutions, with 
$\bar f=\bar b=-2m/r^{d-3}+r^2/\ell^2+2q^2/((d-2)(d-3)r^{2(d-3)})$
for black strings, while  
$\bar f=\bar b=-2m/r^{d-3}+r^2/\ell^2-2q^2/((d-2)(d-3)r^{2(d-3)})$
for  bubble solutions 
(although $\bar a(r)=r^2,~\bar V(r)=q/(r^{d-3}(3-d))+\Phi$ in both cases).
}.
For example,
it is well-known that $\Lambda=0$ uniform black  strings  are 
affected by the Gregory-Laflamme (GL) instability \cite{Gregory:1993vy}.
That is, for a given circle size, the
black objects with translational symmetry bellow a critical
mass are linearly unstable. 
At the critical mass, the marginal mode is static, indicating a new branch of solutions 
with spontaneously broken
translational invariance along the compact direction.

Gregory and Laflamme also noted that the entropy of a fully $d$-dimensional black hole 
is greater than that of the black string with same mass, when $L$ is large enough.
This argument applies also to some of the $\Lambda<0$ vacuum black string solutions,
which
are likely to 
present a GL instability.
This issue is currently under study as well as the issue of
AdS nonuniform black strings, with a dependence on the $z-$coordinate. 

However, charges could prevent the instability of the black 
object since they could contribute some repulsive forces as the horizon shrinks.
Gubser and Mitra \cite{GubserMitra} have proposed an 
interesting conjecture about the relationship between the classical
 black string/brane instability and the local thermodynamic stability. 
This conjecture  asserts that the GL instability 
of the black objects occur iff they are (locally) thermodynamically unstable. 
The solutions discussed in this paper provide a new laboratory
to test this conjecture.
Some of them  are likely to
be stable to linear perturbations.

Another class of $\Lambda<0$ solutions  
expected to exist for the same structure of the
 conformal infinity are black holes
that do not wrap the circle direction.
These solutions would present an
$S^{d-2}$ event horizon topology, their approximate form in 
  the near horizon zone being 
\begin{eqnarray}
\nonumber
ds^2=\frac{d\rho^2}{1-\rho_0^{d-3}/\rho^{d-3} }+\rho^2(d\chi^2+\sin^2 \chi d\Omega_{d-3}^2)
-(1-\rho_0^{d-3}/\rho^{d-3} )dt^2+f_{ij}dx^idx^j~,
\end{eqnarray}
(where $r=\rho \sin \chi$, $z=\rho \cos \chi$ while  $f_{ij}$ are small corrections),
the zeroth order  corresponding to the $d-$dimensional
Schwarzschild metric. 
Their leading order
 asymptotic expansion will be similar to that 
of the black string solutions as given in (\ref{even-inf}), (\ref{odd-inf}).
These would represent the AdS counterparts of 
the $\Lambda=0$ black hole solutions in Kaluza-Klein
theory discussed $e.g.$ in \cite{Kudoh:2004hs}.

We expect the solutions considered in this paper to be relevant in the context of AdS/CFT 
and more generally in the context of gauge/gravity dualities. Indeed, 
as it has been argued at length in \cite{Chamblin:1999tk}, 
physical properties of the RNAdS black holes in $d$
dimensions can be related by means of the AdS/CFT correspondence 
to the physics of a class of $(d-1)$-dimensional field theories that 
are coupled to a global background $U(1)$ current. 
The $k=1$ solutions presented in this paper describe $\Lambda<0$ electrically charged 
and also rotating black strings 
that have horizons with topology $S^1\times S^{d-3}$ and conformal boundary 
$R\times S^1\times S^{d-3}$. According to the AdS/CFT correspondence, their 
properties should be similarly related to those of a dual field theory
living on the AdS boundary. 
In particular, the background metric upon which 
the dual field theory resides is found by taking the rescaling 
$h_{ab}=\lim_{r \rightarrow \infty} \frac{\ell^2}{r^2}\gamma_{ab}$. 
For $k=1$ black strings we find that 
for both charged static and neutral rotating solutions 
the rescaled boundary metric is: 
 \begin{eqnarray}
ds^2=h_{ab}dx^adx^b=-dt^2+dz^2+\ell^2d\Omega_{d-3}^2,
\end{eqnarray}
so that the conformal boundary is indeed $R\times S^1\times S^{d-3}$.

The expectation value of the stress tensor of the dual field theory can be 
computed using the  relation 
\cite{Myers:1999ps}:
\begin{eqnarray} 
\sqrt{-h}h^{ab}<\tau _{bc}>=\lim_{r\rightarrow \infty }\sqrt{-\gamma }\gamma^{ab}T_{bc}.
\end{eqnarray}

Restricting for instance to the case of the electrically charged 
black strings in $5$ dimensions, one can see directly that in the 
asymptotic expansion  (\ref{even-inf}) or (\ref{odd-inf}) of the
 metric functions there is no effect of the electric charge in the 
leading order terms and, therefore, when computing the boundary 
stress-energy tensor one obtains the same components as those 
corresponding to an uncharged black string \cite{Mann:2006yi}. 
It is then apparent that the stress-tensor is finite and 
covariantly conserved, while its trace matches precisely 
the conformal anomaly of the boundary CFT.

Let us consider now the $d=6$ rotating black string solutions 
described in Section $4$. For these solutions we find the 
following non-vanishing components of the boundary 
stress-energy tensor:
\begin{eqnarray}  
\nonumber
&& ~
<\tau^z_z>=-\frac{2c_g+c_t-4 c_z }{16\pi G_6\ell},~
<\tau^{\theta}_{\theta}>=\frac{3c_g-c_t-c_z }{16\pi G_6\ell},~~
<\tau^{\varphi_1}_{\varphi_1}>=\frac{c_g(1+5\cos
2\theta)-2(c_t+c_z)}{32\pi G_6\ell}~,~~{~~} 
\\
\nonumber
&&
<\tau_{\varphi_2}^{\varphi_2}>=\frac{c_g(1-5\cos
2\theta)-2(c_t+c_z)}{32\pi G_6\ell},~
<\tau_{\varphi_1}^{t}>=\frac{c_\varphi \ell \sin^2\theta }{16\pi G_6},~
<\tau^{\varphi_1}_{t}>=<\tau^{\varphi_2}_{t}>=-\frac{5c_g  }{16\pi G_6
\ell},~
\\
\nonumber
&&
<\tau_{\varphi_2}^{t}>=\frac{c_\varphi \ell \cos^2\theta }{16\pi G_6},~
<\tau^{\varphi_2}_{\varphi_1}>=-\frac{5c_g \sin^2\theta }{16\pi G_6
\ell},~
<\tau^{\varphi_1}_{\varphi_2}>=-\frac{5c_g \cos^2\theta }{16\pi G_6
\ell},~
<\tau^{t}_{t}>=\frac{4c_t-c_z-2c_g }{16\pi G_6 \ell}.
\end{eqnarray}
This stress-tensor is traceless, as expected from the absence 
of conformal anomalies for the boundary field theory in odd dimensions. 

More interesting is the situation for the electric black string 
in seven dimensions. First, notice that the 
boundary stress-tensor components are the same with those 
corresponding to the neutral black string. Therefore, we shall 
discuss first the case of the uncharged black string in seven dimensions. 
Indeed, a direct computation of the boundary stress-tensor 
leads to the following components:
\beqs
\nonumber
<\tau^t_t>=\frac{4000c_t-800c_z+63}{12800\pi G_7\ell},~
<\tau^z_z>=\frac{4000c_z-800c_t+63}{12800\pi G_7\ell},~
<\tau^{\theta}_{\theta}>=-\frac{1600c_t+1600c_z+117}{25600\pi G_7\ell},
\eeqs
where by $\theta$ we denote the four angular coordinates on the unit 
sphere $S^4$.

This boundary stress-tensor is finite and covariantly conserved. 
One can also compute its trace and one obtains simply:
 \beqs
 <\tau^{a}_{a}>=-\frac{27}{3200\pi G_7 \ell}.
 \eeqs
As expected from the counterterm Lagrangian, it can 
be checked directly that this trace equals the conformal anomaly \cite{Skenderis:2000in}: 
\beqs
{\cal A}=-\frac{\ell^{5}}{128}\left(\mathsf{RR}^{ab}
\mathsf{R}_{ab}-\frac{3}{25}\mathsf{R}^{3} 
-2\mathsf{R}^{ab}\mathsf{R}^{cd}\mathsf{R}_{acbd}-
\frac{1}{10}\mathsf{R}^{ab}\nabla _{a}\nabla_{b}
\mathsf{R}+\mathsf{R}^{ab}\Box \mathsf{R}_{ab}-\frac{1}{10}\mathsf{R}\Box 
\mathsf{R}\right).
\label{ano1}
\eeqs
According to the AdS/CFT conjecture, this trace anomaly in 
$6$ dimensions should correspond to the trace anomaly of a 
class of $6$-dimensional non-trivial CFT with maximal supersymmetry, 
namely the ${\cal N}=(0,2)$ interacting superconformal theories  
describing $N$ coincident M5 branes \cite{Skenderis:2000in}. 

In general, according to the classification given in \cite{Deser:1993yx} 
one can divide the trace anomalies into Type A, which are proportional 
with the topological Euler density, Type B, which are made-up out of 
Weyl invariants computed using the Weyl tensor, and also, trivial 
anomalies, which can be expressed as total derivatives and can be 
cancelled by the variation of a finite covariant counterterm added to the action. 
Therefore, the trace anomaly can be expressed in $6$ dimensions as 
\cite{Skenderis:2000in}:
\beqs
{\cal A}=-\frac{2a_{(6)}}{16\pi G_7}=-\frac{6\ell^5}{8\pi G_7}
\left(E_{(6)}+I_{(6)}+\nabla_iJ^i\right),
\eeqs
where $E_{(6)}$ is a term proportional to the Euler density 
$E_6$ in $6$ dimensions, $I_{(6)}$ is a Weyl invariant term constructed using 
the $3$ Weyl invariants in $6$ dimensions, while $\nabla_iJ^i$ is a trivial anomaly. 
Before we write the trace anomaly in this form, we shall 
express it in terms of field theory quantities by using the relation
$\ell^5/G_7=16N^3/3\pi^2$ \cite{Skenderis:2000in}. 

Using now the expressions defined in Appendix, we have \cite{Bastianelli:2000hi}:
\beqs
{\cal A}=-\frac{4N^3}{\pi^3}\bigg[\frac{1}{2^{11}3^2}
\left(E_6+8(12I_1+3I_2-I_3)+\frac{2}{35}\nabla_iJ^i\right)\bigg],
\eeqs
where
\beqs
\nabla_iJ^i&=&420C_3-504C_4-840C_5-84C_6+1680C_7.
\eeqs
When evaluated on the boundary geometry corresponding 
to the black string solution, we obtain $E_6=0$, $\nabla_iJ^i=0$,
 while $I_1=-51/100\ell^6$, $I_2=39/25\ell^6$ and $I_3=-684/25\ell^6$.
One can now check explicitly that the above expression 
gives precisely the anomaly (\ref{ano1}). 

It is of interest to compare the above result with that 
corresponding to another maximally supersymmetric CFT in $6$ 
dimensions: namely the non-interacting one, made up by $N$ copies 
of the free ${\cal N}=(0,2)$ tensor multiplet containing five scalars, 
one two-form with selfdual field strength and two Weyl fermions 
\cite{Bastianelli:2000hi,Bastianelli:2000rs}.
The conformal anomaly for the free $(2,0)$ tensor multiplet was 
computed in \cite{Bastianelli:2000hi}:
\beqs
{\cal A}^{free}=-\frac{1}{\pi^3}\bigg[\frac{1}{2^{11}3^2}
\left(\frac{7}{4}E_6+8(12I_1+3I_2-I_3)+\frac{2}{35}\nabla_i{\cal J}^i\right)\bigg],
\eeqs
where
\beqs
\nabla_i{\cal J}^i=56C_2-420C3-420C_5-210C_6+840C_7.
\eeqs
Apart from the expected factor of $4N^3$ \cite{Bastianelli:2000hi}, 
one notices that in the black string background the two anomalies 
are basically the same (since for this background the Euler density 
is identically zero, while $\nabla_i{\cal J}^i=0$ as well). 

As we have previously mentioned, at leading order the components of 
the boundary stress-energy tensor do not exhibit any explicit dependence 
on the electric charge and therefore much of the above discussion 
regarding the conformal anomaly carries through. Since the $7$-dimensional 
gauged supergravities that appear from compactifications of $11$-dimensional 
supergravity do not admit a truncation to pure EM theories with 
negative cosmological constant, it is then apparent that the 
electric black string solution cannot be embedded into any such 
$7$-dimensional gauged supergravity. For this reason the nature 
of the dual boundary field theory 
is unknown. However, in the spirit of AdS/CFT correspondence, the charge carried 
by these solutions can be accounted for by switching on a background global
 $U(1)$ current (or its canonical conjugate charge) to which the dual field theory couples.

As avenues for further research, it might be interesting 
to study the more general case of charged {\it and} rotating AdS 
black string solutions, to which the solutions discussed in 
this paper would be particular cases. However, a more 
complicated action principle than (\ref{action}) is required 
in order to provide a consistent embedding in higher dimensional 
gauged supergravities (in $10$ and $11$ dimensions) of these solutions.
 For example, in $d=5$, the simplest supergravity model contains a
Chern-Simons term. For the static black string solution discussed 
in this paper it is obvious that this term is trivially zero. However, 
it cannot be neglected when discussing general charged rotating
 black string solutions.  
\\
\\
\\
\\
\bigskip
\noindent
{\bf\large Acknowledgements}  \\
The work of ER was carried out in the framework of Enterprise--Ireland Basic Science
Research  Project SC/2003/390 of Enterprise-Ireland.  
 YB is grateful to the
Belgian FNRS for financial support.

\renewcommand{\theequation}{A-\arabic{equation}} 
\setcounter{equation}{0} 

\section*{Appendix}
Following the discussion in \cite{Bastianelli:2000rs,Bastianelli:2000hi} 
we use the following basis of $17$ curvature invariants:
\beqs
\begin{array}{lll}
K_1 = R^3~,  & 
K_2 = R R_{ab}^2~, & 
K_3 = R R_{abmn}^2~, \\ &&\\
K_4 = R_a{}^m R_m{}^i R_i{}^a~, & 
K_5 =  R_{ab} R_{mn} R^{mabn}~, & 
K_6= R_{ab} R^{amnl} R^b{}_{mnl}~, \\ &&\\
K_7 = R_{ab}{}^{mn} R_{mn}{}^{ij}R_{ij}{}^{ab}~, \ \ \ \ &
K_8 = R_{amnb} R^{mijn} R_{i}{}^{ab}{}_j~, \ \ \ \ &
K_9 = R\nabla^2 R~, \\ &&\\
K_{10} = R_{ab}\nabla^2 R^{ab}~, &
K_{11} = R_{abmn}\nabla^2 R^{abmn}~, &
K_{12} = R^{ab} \nabla_a \nabla_b R~, \\&&\\
K_{13} = (\nabla_a R_{mn})^2~,  &
K_{14} = \nabla_a R_{bm} \nabla^b R^{am}~, &
K_{15} = (\nabla_i R_{abmn})^2~,  \\ &&\\
K_{16} = \nabla^2 R^2~, &
K_{17} =\nabla^4 R~. & 
\label{inv}
\end{array}
\eeqs
Then a basis for the non-trivial anomalies in six dimensions can be written as:
\beqs
\label{Ms}
M_{1} &=& \frac{19}{800}K_{1}-\frac{57}{160}K_{2}+\frac{3}{40}K_{3}+
  \frac{7}{16}K_{4} -\frac{9}{8}K_{5}-\frac{3}{4}K_{6}+K_{8},
\nonumber\\
M_{2} &=& \frac{9}{200}K_{1}-\frac{27}{40}K_{2}+\frac{3}{10}K_{3}+
  \frac{5}{4}K_{4}-\frac{3}{2}K_{5}-3K_{6}+K_{7},
\nonumber\\
M_{3} &=& -K_{1}+8K_{2}+2K_{3}-10K_{4}+10K_{5}-\frac{1}{2}K_{9}+5K_{10}-
      5K_{11} ,
\nonumber \\
M_{4} &=& -K_{1}+12K_{2}-3K_{3}-16K_{4}+24K_{5}+24K_{6}-4K_{7}-8K_{8} ,
\nonumber
\\
\nonumber
M_{5} &=& 6K_{6}-3K_{7}+12K_{8}+K_{10}-7K_{11}-11K_{13}+12K_{14} -4K_{15},
\\ 
M_{6} &=& -\frac{1}{5}K_{9}+K_{10}+\frac{2}{5}K_{12}+K_{13},  
\\ 
M_{7} &=& K_{4}+K_{5}-\frac{3}{20}K_{9}+\frac{4}{5}K_{12}+K_{14},  
\nonumber\\
M_{8} &=& -\frac{1}{5}K_{9}+K_{11}+\frac{2}{5}K_{12}+K_{15},  
\nonumber\\
\nonumber
M_{9} &=& K_{16},~~~
M_{10} = K_{17}~. 
\eeqs
From these quantities one can form the following trivial anomalies:
\beqs
&&C_{1}=M_{10}, \ \ \ C_{2}=\frac{1}{2}M_{9}, \ \ \ 
C_{3}=-\frac{1}{12}M_{9}+(K_{10}+K_{13}),
\nonumber\\
&&C_{4}=-\frac{7}{6}M_{6}+M_{7}-\frac{5}{72}M_{9}+\frac{7}{6}(K_{10}+K_{13}), 
 \ \ \ C_{5}=-M_{6}+M_{8}+\frac{1}{20}M_{9},
\\ 
&& C_{6}=2M_{6}-M_{7}+\frac{1}{8}M_{9}-2(K_{10}+K_{13}),
\ \  C_{7}=\frac{1}{12}M_{5}-\frac{1}{12}M_{6}-\frac{1}{4}M_{7}
+\frac{7}{12}M_{8}+\frac{1}{32}M_{9}.
 \nonumber
\eeqs
Finally, one is now ready to compute the topological Euler density in 
$6$ dimensions and the three Weyl invariants \cite{Bastianelli:2000rs}:
\beqs
E_6&=&8M_4~, ~~~~~ I_1=M_1~,~~~~~ I_2=M_2~,
\nonumber\\
I_3&=&\frac{16}{3}M_{1}+\frac{8}{3}M_{2}-\frac{1}{5}M_{3}+
\frac{2}{3}M_{4}-\frac{2}{3}M_{5}-\frac{13}{3}M_{6}+2M_{7}
+\frac{1}{3}M_{8}~.
\eeqs


\end{document}